\documentclass[11pt]{article}
\usepackage{graphicx}

\advance \textheight by \topskip
\makeatletter
\@addtoreset{equation}{section}
    
\makeatother

\def\Z{\mathbb Z}

\def\nn{\nonumber}
\usepackage{amsmath,amssymb}

\newcommand{\be}{\begin{equation}}
\newcommand{\ee}{\end{equation}}
\newcommand{\bea}{\begin{eqnarray}}
\newcommand{\eea}{\end{eqnarray}}
\newcommand{\bi}{\begin{enumerate}}
\newcommand{\ei}{\end{enumerate}}
\newcommand{\bref}[1]{(\ref{#1})}

\newcommand{\T}{\theta} 

\newcommand{\h}{\eta}           
           
\newcommand{\W}{\Omega}

\def\6{\partial}
\def\7{\tilde}
\def\8{\hat}

\def\={{\;=\;}}\def\+{{\;+\;}}

\begin{document}

\title{
{\bf (2+1)D Exotic Newton-Hooke Symmetry,\\ Duality and Projective
Phase}}

\author
{{\sf Pedro D. Alvarez${}^a$}\thanks{E-mail:
pd.alvarez.n@gmail.com}, {\sf Joaquim Gomis${}^b$}\thanks{E-mail:
gomis@ecm.ub.es}, {\sf Kiyoshi Kamimura${}^c$}\thanks{E-mail:
kamimura@ph.sci.toho.u.ac.jp}\, {\sf\ and Mikhail S.
Plyushchay${}^a$}\thanks{E-mail: mplyushc@lauca.usach.cl}
\\[4pt]
{\small \it ${}^a$Departamento de F\'{\i}sica, Universidad de
Santiago de Chile, Casilla 307, Santiago 2, Chile}\\
{\small \it ${}^b$Deparment ECM, Facultat de F\'{\i}sica, Universitat de Barcelona, E-08028, Spain}\\
{\small \it ${}^c$Department of Physics, Toho University Funabashi
274-8510, Japan }}
\date{}

\maketitle

\begin{abstract}
A particle system with a (2+1)D exotic Newton-Hooke symmetry is
constructed by the method of nonlinear realization.  It has three
essentially different phases depending on the values of the two
central charges. The subcritical and supercritical phases
(describing 2D isotropic ordinary and exotic oscillators) are
separated by the critical phase (one-mode oscillator), and are
related by a duality transformation. In the flat limit, the system
transforms into a free Galilean exotic particle on the
noncommutative plane. The wave equations carrying projective
representations of the exotic Newton-Hooke symmetry  are
constructed.
\end{abstract}

\newpage

\section{Introduction}

A revival of the interest to the de Sitter (dS) and anti-de Sitter
(AdS) spacetimes was provoked recently, on the one hand, by
cosmological observations on accelerated expansion of our
universe, and, on the other hand, by investigations on AdS/CFT
correspondence. These  spacetimes of constant curvature are vacuum
solutions of the Einstein equations with   positive and negative
cosmological constants. In a limit when the velocity of light $c$
tends to infinity while cosmological constant $\Lambda$ tends to
zero but keeping $c^2\Lambda$ finite, the symmetries of dS and AdS
spacetimes become what are called the Newton-Hooke (NH)
symmetries, which  in a flat limit reduce to the ordinary Galilei
symmetry \cite{BacLev,BacNuy,AldBar,Mariano,Gao,GibPat,BGK}.

In a general case of D dimensions the Galilei and NH groups admit an
extension with a one central charge, while in a special case of 2+1
dimensions, they admit an exotic, two-fold central extension
\cite{LL,BGO,BGGK,Mariano}. The central extensions are  related to a
nontrivial Eilenberg-Chevalley  cohomology of the corresponding Lie
algebras. The exotic (2+1)D Galilei symmetry recently has attracted
the attention in the context of research on non-commutative geometry
and condensed matter physics
\cite{LSZ,DH1,JackN,DH2,HP1,PHMP,HP3,mariano1,OP,Horv}. It is a
symmetry of a free relativistic particle in a non-commutative plane,
which appears from a (2+1)D theory of relativistic anyons in a
special non-relativistic limit \cite{JackN,HP1,PHMP}. Surprisingly,
until now, the exotic Newton-Hooke symmetry has not attracted much
attention.

The present paper is devoted to investigation of  classical and
quantum aspects of the (2+1)D  exotic Newton-Hooke (NH) symmetry.
We start by  constructing the exotic NH$_3$ algebra as a
contraction of AdS$_3$ algebra. We then construct  a particle
system with the (2+1)D exotic NH symmetry  by the method of
nonlinear realization, and analyze its dynamics and constraint
structure. We quantize the system in a reduced phase space, and
show that the model has three essentially different phases
depending on the values of the two central charges. The
subcritical and supercritical phases describe 2D isotropic
ordinary and exotic oscillators, whose spectra are characterized,
respectively,  by additional SO(3) rotational and SO(2,1) Lorentz
symmetry. These two phases  are separated by the critical phase
corresponding to a one-mode oscillator,  and are related by a
duality transformation. We demonstrate that in the flat limit the
system transforms into a free Galilean exotic particle on the
noncommutative plane. For the obtained system we construct wave
equations carrying projective representations of the exotic
Newton-Hooke  symmetry. We find the projective phase related to
the nontrivial cohomology of the exotic NH symmetry, which, being
associated with quasi invariance of the Lagrangian under finite NH
transformations, guarantees invariance of the wave equations. We
also discuss shortly some possible applications and
generalizations of the exotic Newton-Hooke symmetry.

The paper is organized as follows. In Section \ref{ADS3}, the exotic
NH$_3$ algebra is obtained by a contraction of AdS$_3$. In Section
\ref{classicalLag} various forms of a classical Lagrangian for a
particle with exotic NH$_3$ symmetry are constructed. In Sections
\ref{classicaldyn}, \ref{clasconstraints} and \ref{classicalNH} we
analyze classical dynamics of the system, its constraints,
realization of exotic Newton-Hooke symmetry, and a manifest
additional  symmetry depending on the phase.  In Section
\ref{reducedquant} we quantize the system in the reduced phase
space. In Section \ref{addsymduality} we discuss duality
transformations. Wave equations for the system are constructed in
Section \ref{waveEquat}, where we also discuss the flat limit. In
Section \ref{Phaseproj} we compute the projective phase related to
the nontrivial cohomology. Section \ref{discussionoutlook} is
devoted to the discussion and outlook. In Appendix  we compute a
non-trivial Eilenberg-Chevalley cohomology for the Newton-Hooke
group in 2+1 dimensions.

\section{Exotic NH${}_3$ as a contraction of
AdS${}_3$}\label{ADS3}

The AdS${}_3$ algebra is given by
\begin{eqnarray}
    &[M_{\mu\nu},M_{\lambda\sigma}]=i\left(
     \eta_{\mu\lambda}M_{\nu\sigma}-\eta_{\mu\sigma}M_{\nu\lambda}+
     \eta_{\nu\sigma}M_{\mu\lambda}-\eta_{\nu\lambda}M_{\mu\sigma}\right),\label{AdS1}\\
    &[M_{\mu\nu},P_{\lambda}]=i\left(\eta_{\mu\lambda}P_\nu
    -\eta_{\nu\lambda}P_\mu\right),&\label{AdS2}\\
    &[P_\mu,P_\nu]=-i\frac{1}{R^2}M_{\mu\nu},&\label{AdS3}
\end{eqnarray}
where $\mu=0,1,2$, $\eta_{\mu\nu}=diag(-,+,+)$. If we define
$M_{\mu 3}=RP_\mu$, the AdS${}_3$ algebra takes a form of the
$SO(2,2)$ algebra. To perform the non-relativistic contraction,
rewrite (\ref{AdS1})--(\ref{AdS3}) in the decomposed form
($\mu=0,i$; $i,j=1,2$, $\epsilon_{12}=1$),
\begin{equation}\label{P0M12}
    [P_0,M_{12}]=0,
\end{equation}
\begin{equation}\label{P0M0i}
    [P_0,M_{0i}]=iP_i,\qquad
    [P_0,P_i]=-i\frac{1}{R^2}M_{0i},
\end{equation}
\begin{equation}\label{M12Pi}
    [M_{12},P_i]=i\epsilon_{ij}P_j,\qquad
    [M_{12},M_{0i}]=i\epsilon_{ij}M_{0j},
\end{equation}
\begin{equation}\label{M0iPj}
    [M_{0i},P_j]=-i\delta_{ij}P_0,\qquad
    [M_{0i},M_{0j}]=-i\epsilon_{ij} M_{12},
\end{equation}
\begin{equation}\label{PiPj}
    [P_i,P_j]=-i\frac{1}{R^2}\epsilon_{ij}M_{12},
\end{equation}
and then replace the radius $R$ and the AdS${}_3$ generators for the
rescaled quantities,
\begin{equation}\label{contr}
    P^0=-P_0\rightarrow \omega Z +\frac{1}\omega H,\qquad
    M_{0i}\rightarrow  \omega K_i,\qquad
    M_{12}\rightarrow  \omega^2\tilde{Z}+J,\qquad
    R\rightarrow \omega R.
\end{equation}
We introduce two new generators $Z$ and $\tilde{Z}$ that have a
nature of central charges because the unextended NH${}_3$ algebra
has a non-trivial Chevalley-Eilenberg cohomology group for
differential forms of degree two, see Appendix. Taking a limit
$\omega\rightarrow\infty$, we get the exotic NH${}_{3}$ algebra with
two central charges,
\begin{equation}\label{NH3ex}
    [H,J]=0,
\end{equation}
\begin{equation}\label{HPK}
    [H,K_i]=-iP_i,\qquad [H,P_i]=i\frac{1}{R^2}K_i,
\end{equation}
\begin{equation}\label{JKP}
    [J,P_i]=i\epsilon_{ij}P_j,\qquad
    [J,K_i]=i\epsilon_{ij}K_j,
\end{equation}
\begin{equation}\label{KPK}
    [K_i,P_j]=i\delta_{ij}Z,\qquad
    [K_i,K_j]=-i\epsilon_{ij}\tilde{Z},
\end{equation}
\begin{equation}\label{PP}
    [P_i,P_j]=-i\frac{1}{{R}^2}\epsilon_{ij}\tilde{Z}.
\end{equation}
This algebra was considered before in \cite{Mariano,Gao}. In the
flat limit $R\rightarrow \infty$ it transforms into algebra of
exotic Galilei symmetry \cite{LL,BGO,BGGK}.

The quadratic Casimirs of NH${}_3$ can be obtained from the two
AdS${}_3$ quadratic Casimirs
\begin{equation}\label{CasAdS1}
    C_1=-P_\mu
    P^\mu+\frac{1}{2R^2}M_{\mu\nu}M^{\mu\nu}=P_0^2-P_1^2-P_2^2+\frac{1}{R^2}
    (J_0^2-J_1^2-J_2^2),
\end{equation}
\begin{equation}\label{CasAdS2}
    C_2=-P_\mu J^\mu=P_0J_0-P_1J_1-P_2J_2,
\end{equation}
corresponding to the two SO(2,2) Casimirs $M_{mn}M^{mn}$ and
$\epsilon_{mnls}M^{mn}M^{ls}$, where
\begin{equation}
    J_{\mu}=\frac{1}{2}\epsilon_{\mu\nu\lambda}M^{\nu\lambda}
\label{Jmu}
\end{equation}
 with $\epsilon_{012}=+1$. Note that
we should consider  $C_1$ and $\frac{1}{R}C_2$ as Casimirs of the
same dimension. If we perform the contraction (\ref{contr}), the
finite part of the expansion gives
\begin{equation}\label{CasNH1}
    {\cal C}_1=2\left(ZH+\frac{1}{{R}^2}\tilde{Z}J\right)
    -P_i^2-\frac{1}{{R}^2}K_i^2,
\end{equation}
\begin{equation}\label{CasNH2}
    {\cal C}_2=-ZJ-\tilde{Z}H-\epsilon_{ij}P_iK_j.
\end{equation}

It is convenient also to present the NH${}_3$ in another, chiral
basis. It's existence is rooted in the algebra isomorphism
\begin{equation}\label{Isomor}
    AdS_3\sim SO(2,2)\sim SO(2,1)\oplus SO(2,1)\sim
    AdS_2\oplus AdS_2.
\end{equation}
Defining
\begin{equation}\label{Jpmdef}
J^\pm_\mu=\frac{1}{2}(J_\mu\mp RP_\mu),
\end{equation}
we get the equivalent form of AdS${}_3$ algebra,
\begin{equation}\label{JJAdS}
    [J^\pm_\mu,J^\pm_\nu]=i\epsilon_{\mu\nu}{}^\lambda
    J^\pm_\lambda,\qquad
    [J^+_\mu,J^-_\nu]=0,
\end{equation}
with two Casimirs
\begin{equation}\label{Cas+-}
    C^\pm=\eta^{\mu\nu}J^\pm_\mu
    J^\pm_\nu={J^\pm_1}^2+{J^\pm_2}^2-{J^\pm_0}^2.
\end{equation}
Note that here $R$ is hidden in  definition (\ref{Jpmdef}).
 In correspondence with the non-relativistic contraction
realized in the non-chiral basis, we replace
\begin{equation}\label{contrchiral}
    J^\pm_0\rightarrow -\omega^2Z^\pm +{\cal J}^\pm,\qquad
    J^\pm_i\rightarrow \omega {\cal J}^\pm_i,
\end{equation}
and take the limit $\omega\rightarrow \infty$. As a result we
obtain the exotic NH${}_3$ algebra in the chiral basis
\begin{equation}\label{NHchiral}
    [{\cal J}^\pm_i,{\cal J}^\pm_j]=i\epsilon_{ij}Z^\pm,\qquad
    [{\cal J}^\pm,{\cal J}^\pm_i]=i\epsilon_{ij}{\cal J}^\pm_j,\qquad
    [G^+_a,G^-_b]=0,
\end{equation}
where $G^\pm_a=Z^\pm,{\cal J}^\pm,{\cal J}^\pm_i$. The quadratic
Casimirs are
\begin{equation}\label{CasNHch}
    {\cal C}^\pm={{\cal J}_i^\pm}^2+2Z^\pm{\cal J}^\pm.
\end{equation}
In correspondence with decomposition (\ref{Isomor}), we find that
two-fold centrally extended NH${}_3$ algebra is a direct sum of
two centrally extended NH${}_2$ algebras  obtained by a
contraction of corresponding AdS${}_2$ chiral components. Note
that usual Newton-Hooke algebra in d dimensions NH${}_d$ has only
one central extension, $Z$.

The relation between generators in the chiral and non-chiral bases
is
\begin{equation}\label{NHchNch}
    {\cal J}^\pm=\frac{1}{2}\left(J\pm  R H\right),\qquad
    {\cal J}^\pm_i=\frac{1}{2}\left(\epsilon_{ij}K_j\mp  RP_i\right),\qquad
    Z^\pm=-\frac{1}{2}\left(\tilde{Z}\pm  RZ\right).
\end{equation}

\section{Classical Lagrangian for a system with exotic NH${}_3$
symmetry}\label{classicalLag}

To construct a Lagrangian for the exotic NH particle by the method
of non-linear realization  \cite{Coleman}, we should choose a
suitable coset $\frac{\cal G}{\cal H}$. In our case ${\cal G}$ is
the double extended NH$_{3}$ and ${\cal H}$ is  the rotation group
in two dimensions. Locally we can parametrize the elements of the
coset as
\begin{equation}
\label{coset1} g=e^{-iH x^0}\;e^{iP_ix^i}\;e^{iK_jv^j} \;e^{iZc}
\;e^{-i\tilde Z\tilde c}.
\end{equation}
The (Goldstone)  coordinates of the coset depend on the parameter
$\tau$ that parametrizes the world line of the particle, see for
example  \cite{West,GomisKW}.
 The Maurer-Cartan one-form is
 \begin{equation}\label{MCar}
    \Omega=-ig^{-1}dg= -L_H H +L^i_P P_i+L^i_K K_i+L_J J
    -L_Z Z+ L_{\tilde Z} \tilde Z\, ,
\end{equation}
where
\begin{eqnarray}
    L_H &=& dx^0,
    \qquad
    L^i_P = dx^i-v^i dx^0,
    \qquad
    L^i_K = dv^i+\frac{x^i}{R^2}\,dx^0,
    \qquad
    L_J = 0, \nn\\
    L_Z &=& -dc-v^idx^i+\frac{v_i^2}{2}dx^0+\frac{x_i^2}{2R^2}\,dx^0,
    \nn\\
    L^{\tilde Z} &=& -d\tilde c -\frac{1}{2}\epsilon_{ij}
    \left(v^idv^j+\frac{1}{R^2}x^i dx^j-
    \frac{2}{R^2}x^iv^jdx^0\right). \label{MCNH2}
\end{eqnarray}
The one-form (\ref{MCar}) satisfies the Maurer-Cartan equation
$d\Omega+i\Omega\wedge\Omega=0$.

The Lagrangian is obtained as a pullback of a  linear combination
of nonzero one-forms invariant under rotations, $L_H, L_Z,
L_{\tilde Z}$,
\begin{equation}\label{nhactionZZ}
{\cal L}=\mu\dot{x}^0+
    m\left(\dot c+v^i\dot x^i-\frac{v_i^2}{2}\dot{x}^0-\frac{x_i^2}{2R^2}\dot{x}^0\right)\\
    +\kappa\left(\dot{\tilde c}+\frac{1}{2}\epsilon_{ij}\left(v^i\dot v^j+\frac{1}{R^2}x^i \dot
    x^j- \frac{2}{R^2}x^iv^j\dot{x}^0\right)\right),
\end{equation}
where $\mu$ and $m$  are two parameters of dimension
 $R^{-1}$ while parameter $\kappa$ is dimensionless.
If further fix the diffeomorphism invariance by choosing
$\tau=x^0\equiv t$ and discard total derivative terms, we get
\begin{equation}\label{nhaction}
{\cal L}_{nc}=
    m\left(v_i\dot x_i-\frac{v_i^2}{2}-\frac{x_i^2}{2R^2}\right)\\
    +\frac 12\kappa\epsilon_{ij}\left(v_i\dot v_j+\frac{1}{R^2}x_i \dot
    x_j- \frac{2}{R^2}x_iv_j\right).
\end{equation}

For the analysis of the dynamics and subsequent quantization of the
system, it is more convenient to use another form of Lagrangian
which can be obtained via a simple change of variables and
parameters
\begin{eqnarray}\label{Xpm}
    X^\pm_i&=&x_i\pm
    R\epsilon_{ij}v_j,\label{Xpm*}\\
    \mu_\pm&=&\frac{1}{2R^2}(mR\pm
    \kappa).\label{Mpm}
\end{eqnarray}
In their terms Lagrangian (\ref{nhaction}) rewrites
\begin{equation}\label{Lchir}
    {\cal L}_{ch}={\cal L}_++{\cal L}_-=-\frac{1}{2}
    \mu_+\left(\epsilon_{ij}\dot{X}^+_iX^+_j+\frac{1}{R}{X^+_i}^2\right)
    -\frac{1}{2}
    \mu_-\left(-\epsilon_{ij}\dot{X}^-_iX^-_j+\frac{1}{R}{X^-_i}^2\right).
\end{equation}
Coordinates $X^\pm_i$ have a sense of chiral modes of the exotic NH
particle. Non-chiral, ${\cal L}_{nc}$, and chiral, ${\cal L}_{ch}$,
forms of Lagrangian coincide up to a total time derivative term,
\begin{equation}\label{Nc-Ch}
    {\cal L}_{nc}={\cal L}_{ch}+\frac{d}{dt}\Delta {\cal L},\qquad
    \Delta{\cal
    L}=\frac{m}{2}x_iv_i=\frac{m}{4R}\epsilon_{ij}X^+_iX^-_j,
\end{equation}
which has been omitted when we passed from  (\ref{nhaction}) to
(\ref{Lchir}), but will be important in quantum theory.

Lagrangian (\ref{Lchir}) can also be obtained by the method of
nonlinear realization from the chiral form of exotic NH$_3$ algebra
(\ref{NHchiral}). In this case the elements of the coset are
parametrized as
\begin{equation}\label{cosetchir}
    g=g^+g^-,
\end{equation}
with
\begin{eqnarray}\label{g+-}
    g^+ &=& e^{-ix^0R^{-1}{\cal J}^+}e^{-iR^{-1}X^-_j{\cal J}^+_j}e^{ic^+Z^+}, \nn\\
    g^- &=& e^{ix^0R^{-1}{\cal
    J}^-}e^{iR^{-1}X^+_j{\cal J}^-_j}e^{ic^-Z^-}.
\end{eqnarray}
 Note that there is only one time evolution variable $x^0$ for
these chiral sectors. Difference in signs in exponents with $x^0$
originates from the relation $H=({\cal J}^+-{\cal J}^-)/R$.

\section{Classical dynamics}\label{classicaldyn}

Variation of Lagrangian (\ref{nhaction}) in $v_i$ and $x_i$ gives
the equations of motion
\begin{eqnarray}
&&m(\dot{x}_i-v_i)+{\kappa}\epsilon_{ij}\left(\dot{v}_j+R^{-2}x_j\right)=0,
\label{eom1}
\\
&&m\left(\dot{v}_i+R^{-2}
x_i\right)-R^{-2}\epsilon_{ij}(\dot{x}_j-v_j)=0. \label{eom2}
\end{eqnarray}
They can be presented in an equivalent  form
\begin{eqnarray}
&&(\dot{x}_i-v_i)+\rho
R\epsilon_{ij}\left(\dot{v}_j+R^{-2}x_j\right)=0, \label{eom1'}
\\
&&(\dot{x}_i-v_i)+\rho^{-1}
R\epsilon_{ij}\left(\dot{v}_j+R^{-2}x_j\right)=0, \label{eom2'}
\end{eqnarray}
where
\begin{equation}  \label{rho}
\rho=\frac{\kappa}{mR}.
\end{equation}
For $\rho^2\neq 1$, the parameter transformation
\begin{equation}\label{rhosym}
    \rho\rightarrow \rho^{-1}
\end{equation}
provokes a mutual change of equations (\ref{eom1'}) and
(\ref{eom2'}). These equations   imply
\begin{equation}\label{vdotx}
    \dot x_i-v_i=\dot{v}_i+R^{-2}x_i=0
    \qquad\to \qquad \ddot x_i+\frac{1}{R^2}x_i=0.
\end{equation}
The system looks like a planar isotropic oscillator in both cases
$\rho^2<1$ and $\rho^2>1$. We shall see that in the latter case,
however, it is characterized by some exotic properties.

At critical values  $\rho=\varepsilon$, $\varepsilon=\pm 1$, which
separate subcritical, $\rho^2<1$, and supercritical, $\rho^2>1$,
phases, Eqs. (\ref{eom1'}), (\ref{eom2'}) reduce to only one vector
equation
\begin{equation}\label{eqrhocrit}
    (\dot{x}_i-v_i)+{\varepsilon}\,
    R\epsilon_{ij}(\dot{v}_j+R^{-2}x_j)=0.
\end{equation}
This reflects a gauge symmetry appearing in the system in the
critical phase $\rho^2=1$,
\begin{equation}\label{gauge}
    \delta_\sigma x_i=\sigma_i(t),\quad
    \delta_\sigma v_i=
    {\varepsilon}R^{-1}\epsilon_{ij}\sigma_j(t).
\end{equation}

The picture becomes more transparent if we consider the dynamics
in terms of  chiral  coordinates (\ref{Xpm}). In the noncritical
case transformation (\ref{rhosym}) assumes a change
$mR\leftrightarrow \kappa$, that corresponds to a transformation
\begin{equation}\label{M+M-}
    \mu_\pm\rightarrow \pm \mu_\pm.
\end{equation}
 Lagrangian (\ref{Lchir}) is invariant under transformation
 (\ref{M+M-})  accompanied with a complex coordinate transformation
\begin{equation}\label{X+X-trans}
    X^+_i\rightarrow X^+_i,\qquad X^-_i\rightarrow iX^-_i.
\end{equation}
The appearance of the pure imaginary factor $i$ in transformation
(\ref{X+X-trans}) is related to some peculiar properties of the
supercritical phase $\rho^2>1$ to be discussed below. Here it is
worth noting, however,  that transformation (\ref{M+M-}),
(\ref{X+X-trans}) is not a usual (discrete) symmetry of the system
since it involves the change of the parameters. Moreover,
transformation (\ref{X+X-trans}) itself belongs to a broad class
of similarity transformations $X^+_i\rightarrow \gamma^+ X^+_i$,
$X^-_i\rightarrow \gamma^- X^-_i$, where $\gamma^\pm$ are
numerical parameters, which leave invariant equations of motion
but change Lagrangian. Such transformations cannot be promoted to
symmetries at the quantum level, see Ref. \cite{Gozzi} for the
discussion.

For $\rho^2\neq 1$ equations of motion generated by Lagrangian
(\ref{Lchir}) are
\begin{equation}\label{X+-Eq}
    \dot{X}^\pm_i\pm \frac{1}{R}\epsilon_{ij}X^\pm_j=0.
\end{equation}
The coordinates $X^+_i$ and $X^-_i$ have a sense of normal modes
realizing `right' and `left' rotations of the same frequency
$R^{-1}$,
\begin{equation}\label{X+-motion}
    X^\pm_i(t)=X^\pm_i(0)\cos R^{-1}t\mp \epsilon_{ij}X^\pm_j(0)\sin
    R^{-1}t.
\end{equation}
The evolution of $x_i$ and $v_i$ is a linear superposition of these
two rotations,
$$
x_i=\frac{1}{2}(X^+_i+X^-_i),\qquad
v_i=\frac{1}{2R}\epsilon_{ij}(X^-_j-X^+_j).
$$

In the critical case $\rho=1$ ($\rho=-1$) in accordance with Eq.
(\ref{Mpm}) we have $\mu_-=0$ ($\mu_+=0$). The chiral coordinates
$X^-_i$ ($X^+_i$) disappear from Lagrangian (\ref{Lchir})
transforming into pure gauge variables. This corresponds to a
reduction of equations  (\ref{eqrhocrit}) in the non-chiral basis
at $\rho=\varepsilon$.

Eq. (\ref{X+-Eq}) obtained by variation of (\ref{Lchir}) in
$X^\pm_i$, can be  rewritten as $X^\pm_i\mp R\epsilon_{ij}\dot
X^\pm_j=0$. The coordinate $X^\pm_i$ can be eliminated in terms of
velocity of $X^\pm_j, j\neq i$. Therefore any fixed component of the
vectors $X^+_i$ and $X^-_i$ can be treated as an auxiliary variable.
If we choose $i=2$ for both chiral modes, substitute
\begin{equation}
X^\pm_2=\mp R\dot X^\pm_1 \label{X2X1dot}
\end{equation}
into Lagrangian, and  omit the total derivative term, we get an
equivalent form for the Lagrangian,
\begin{equation}\label{Lshift}
    {\cal L}=
    \frac{1}{2}\mu_+R\left((\dot{X}^+_1)^2 -\frac{1}{R^2}X^{+2}_1\right)
    +\frac{1}{2}\mu_-R
    \left((\dot{X}^-_1)^2-\frac{1}{R^2}X^{-2}_1\right).
\end{equation}
This form of Lagrangian shows explicitly that the system is a sum of
two  harmonic oscillators of the same frequency, but it hides a 2D
rotation invariance of the system.

It is worth noting that the equation $v_i=\dot{x}_i$ is generated by
variation of Lagrangian (\ref{nhaction}) in some linear combination
of $v_i$ and $x_i$, but not in $v_i$ itself. If we try to substitute
this equation into (\ref{nhaction}), as a consequence we get a
higher derivative Lagrangian for $x_i$, which will produce equations
of motion for $x_i$ to be not equivalent to those in (\ref{vdotx})
\cite{HeTe}.

\section{Constraints}\label{clasconstraints}

\subsection{Non-chiral basis}

System (\ref{nhaction}) has two pairs of primary constraints
\begin{eqnarray}\label{constNon1}
    \Pi_i &=& \pi_i+\frac{\kappa}{2}\epsilon_{ij}v_j\approx 0,\\
    V_i &=& p_i-mv_i+\frac{\kappa}{2R^2}\epsilon_{ij}x_j\approx
    0,\label{constNon2}
\end{eqnarray}
where $p_i$ and $\pi_i$ are the momenta canonically conjugate to
$x^i$ and $v^i$. Constraints (\ref{constNon1}) satisfy relations
\begin{equation}\label{VPialg}
    \{\Pi_i,\Pi_j\}=\kappa\epsilon_{ij},\qquad
    \{V_i,V_j\}=\frac{\kappa}{R^2}\epsilon_{ij},\qquad
    \{\Pi_i,V_j\}=m\delta_{ij}.
\end{equation}
Determinant of the matrix of Poisson brackets of the constraints,
$A_{ab}=\{\phi_a,\phi_b\}$, $\phi_a=(\Pi_i,V_i)$,
 is
\begin{equation}\label{detA}
    \det A= m^4(1-\rho^2)^2.
\end{equation}
In non-critical case the matrix is non-degenerate, and
(\ref{constNon1}), (\ref{constNon2}) form the set of second class
constraints. At $\rho^2=1$ the matrix degenerates, there are first
class constraints which we will analyze separately.

Canonical Hamiltonian corresponding to non-chiral Lagrangian
(\ref{nhaction}) is
\begin{equation}\label{HcanNC}
    H_{can}=\frac{m}{2}\left(v_i^2+\frac{1}{R^2}x_i^2\right)+\frac{\kappa}{R^2}\epsilon_{ij}
    x_iv_j.
\end{equation}
Adding to it a linear combination of primary constraints and
applying the Dirac algorithm, for the non-critical case we get a
total Hamiltonian
\begin{equation}\label{HtotNC}
    H=p_iv_i-\frac{\pi_i x_i}{R^2}+\frac{m}{2}\left(
    \frac{x_i^2}{R^2}-v_i^2\right),
\end{equation}
\begin{equation}\label{NCconsistency}
    \{H,\Pi_i\}=V_i,\qquad
    \{H,V_i\}=-\frac{1}{R^2}\Pi_i.
\end{equation}

The following linear combinations of the phase space coordinates,
\begin{equation}\label{obserNC}
    {\cal P}_i=p_i-\frac{\kappa}{2R^2}\epsilon_{ij}x_j,\qquad
    {\cal
    K}_i=mx_i-\pi_i+\frac{1}{2}\kappa\epsilon_{ij}v_j,
\end{equation}
have zero Poisson brackets with constraints (\ref{constNon1}) and
(\ref{constNon2}). These are observable variables which satisfy
relations
\begin{equation}\label{PPKK}
    \{{\cal P}_i,{\cal P}_j\}=-\frac{\kappa}{R^2}\epsilon_{ij},\qquad
    \{{\cal K}_i,{\cal P}_j\}=m\delta_{ij},\qquad
    \{{\cal K}_i,{\cal K}_j\}=-\kappa\epsilon_{ij}.
\end{equation}

For $\kappa\neq 0$ constraints (\ref{constNon1})  form a subset of
second class constraints. Reduction on their surface allows us to
express $\pi_i$ in terms of $v_i$, and results in nontrivial
Poisson-Dirac brackets
\begin{equation}\label{xpvv}
    \{x_i,p_j\}=\delta_{ij},\qquad
    \{v_i,v_j\}=-\frac{1}{\kappa}\epsilon_{ij}.
\end{equation}
In terms of these brackets,
\begin{equation}\label{VV}
    \{V_i,V_j\}=-\frac{m^2}{\kappa}(1-\rho^2)\epsilon_{ij}.
\end{equation}
In correspondence with (\ref{detA}), for $\rho^2\neq 1$
constraints (\ref{constNon2}) are second class. At $\rho^2=1$ they
are first class, and dimension of a physical subspace is less in
two in comparison with a noncritical case. For $\rho^2\neq 1$,
subsequent reduction to the surface of second class constraints
(\ref{constNon2}) excludes the variables $v_i$, and for
independent reduced phase space variables $x_i$ and $p_i$ we get
\begin{equation}\label{xxpx}
    \{x_i,x_j\}=\frac{\kappa}{m^2}\,\frac{1}{1-\rho^2}\epsilon_{ij},\qquad
    \{x_i,p_j\}=\frac{1-\frac{1}{2}\rho^2}{1-\rho^2}\delta_{ij},\qquad
    \{p_i,p_j\}=\frac{m^2}{4\kappa}\,\frac{\rho^4}{1-\rho^2}\epsilon_{ij}.
\end{equation}
 Dynamics is generated by the reduced
phase space Hamiltonian
\begin{equation}\label{Hredxp}
    H^{*}=\frac{1}{2m}\left(p_i-\frac{\kappa}{2R^2}\epsilon_{ij}x_j\right)^2+
    \frac{m}{2R^2}(1-\rho^2)x_i^2
\end{equation}
being a restriction of the total Hamiltonian (\ref{HtotNC}), to
the surface of the constraints (\ref{constNon1}),
(\ref{constNon2}).

{}From the explicit form of Hamiltonian (\ref{Hredxp}) it is clear
that the subcritical, $\rho^2<1$, and supercritical, $\rho^2>1$,
cases are essentially different: in the first case energy has a sign
of a parameter $m$, while in the second case it can take both signs.

Note that the symplectic structure  of the reduced phase space
(\ref{xxpx}) has a form similar to that for the Landau problem in
the noncommutative plane \cite{DH1,HP3}.
 In the flat limit $R\rightarrow
\infty$, the symplectic structure and Hamiltonian (\ref{Hredxp})
take the form of those for a free particle on the noncommutative
plane, which is described by the exotic Galilean symmetry with
non-commuting boosts \cite{PHMP}.

\subsection{Chiral basis}

In the chiral basis, the system with Lagrangian (\ref{Lchir}) is
described by the set of constraints
\begin{eqnarray}\label{chi+}
    &\chi^+_i = P^+_i+\frac{1}{2}\mu_+\epsilon_{ij}X^+_j\approx
    0,&\\
    &\chi^-_i = P^-_i-\frac{1}{2}\mu_-\epsilon_{ij}X^-_j\approx
    0.&\label{chi-}
\end{eqnarray}
Here $P^\pm_i$ are the momenta canonically conjugate to $X^\pm_i$,
\begin{equation}\label{Pppi}
    P^\pm_i=\frac{1}{2}\left(p^{ch}_i\pm
    \frac{1}{R}\epsilon_{ij}\pi^{ch}_j\right).
\end{equation}
The momenta $p^{ch}_i$ and $\pi^{ch}_i$ canonically conjugate to
$x_i$ and $v_i$ are related to the corresponding momenta for
non-chiral Lagrangian (\ref{nhaction}) by the canonical
transformation
\begin{equation}\label{chi-nochi}
    p_i^{ch}=p_i-\frac{m}{2}v_i,\qquad
    \pi^{ch}_i=\pi_i-\frac{m}{2}x_i
\end{equation}
associated with time derivative shift (\ref{Nc-Ch}).

Nonzero brackets of the constraints are
\begin{eqnarray}\label{mu+}
    \{\chi^+_i,\chi^+_j\}&=&\mu_+\epsilon_{ij},\\
    \{\chi^-_i,\chi^-_j\}&=&-\mu_-\epsilon_{ij}.\label{mu-}
\end{eqnarray}
In the noncritical case $\rho^2\neq 1$, these are two sets of second
class constraints. When $\mu_+=0$ ($\mu_-=0$), the constraints
$\chi^+_i$ ($\chi^-_i$) change their nature from the second to the
first class,  taking a form $\chi^+_i=P^+_i\approx 0$
($\chi^-_i=P^-_i\approx 0$). In this case the degrees of freedom
corresponding to the `$+$' (`$-$') mode are pure gauge, that
reflects disappearance of the corresponding term ${\cal L}_+$
(${\cal L}_-$) from Lagrangian (\ref{Lchir}).

Canonical Hamiltonian corresponding to the chiral Lagrangian
(\ref{Lchir}) is
\begin{equation}\label{H+-}
    H_{can}=\frac{1}{2R}\left(\mu_+{X^+_i}^2+\mu_-{X^-_i}^2\right).
\end{equation}
Define a total Hamiltonian
$H=H_{can}+u^+_i\chi^+_i+u^-_i\chi^-_i$. A requirement of
conservation of constraints fixes multipliers, $u^\pm_i=\mp
\frac{1}{R}\epsilon_{ij}X^\pm_j$, and gives
\begin{equation}\label{Htotchir}
    H=\frac{1}{R}\epsilon_{ij}(X^+_iP^+_j-X^-_iP^-_j).
\end{equation}
This Hamiltonian reproduces Lagrangian equations of motion for
chiral coordinates (\ref{X+-Eq}).

Independent variables  (observables) commuting with constraints are
\begin{equation}\label{lambdai}
    \lambda^+_i= P^+_i-\frac{1}{2}\mu_+\epsilon_{ij}X^+_j,\qquad
    \lambda^-_i = P^-_i+\frac{1}{2}\mu_-\epsilon_{ij}X^-_j,
\end{equation}
\begin{equation}\label{lamchi}
    \{\lambda^+_i,\chi^\pm_j\}=\{\lambda^-_i,\chi^\pm_j\}=0.
\end{equation}
They have brackets similar to the brackets of the constraints,
\begin{equation}\label{lamlam}
    \{\lambda^+_i,\lambda^+_j\}=-\mu_+\epsilon_{ij},\qquad
    \{\lambda^-_i,\lambda^-_j\}=\mu_-\epsilon_{ij},\qquad
    \{\lambda^+_i,\lambda^-_j\}=0.
\end{equation}

Reduction to the surface given by (\ref{chi+}) (when $\mu_+\neq 0$)
and/or (\ref{chi-}) (when $\mu_-\neq 0$) results in exclusion of the
momenta $P^+_i$ and/or $P^-_i$.  Nontrivial Dirac brackets are
\begin{eqnarray}\label{XX++}
    \{X^+_i,X^+_j\} & =& -\frac{1}{\mu_+}\epsilon_{ij},\\
    \{X^-_i,X^-_j\} & =& \frac{1}{\mu_-}\epsilon_{ij},\label{XX--}
\end{eqnarray}
and reduced phase space Hamiltonian
\begin{equation}\label{H+-*}
    H^*=\frac{1}{2R}\left(\mu_+{X^+_i}^2+\mu_-{X^-_i}^2\right)
\end{equation}
coincides with the form of canonical Hamiltonian (\ref{H+-}).  In
correspondence with non-chiral picture, for subcritical case
$\rho^2<1$, $\mu_\pm>0$ for $m>0$ and $\mu_\pm<0$  for $m<0$, and
energy takes values of the sign of the parameter $m$. In the
supercritical case $\rho^2>1$ the constants $\mu_+$ and $\mu_-$ have
opposite signs, and energy can take values of any sign.

\section{Symmetries}\label{classicalNH}
\subsection{Classical exotic Newton-Hooke symmetry}\label{classicalNH1}

Lagrangian (\ref{nhaction}) is quasi-invariant under transformations
generalizing Galilean translations and boosts,
\begin{eqnarray}
&x_i^{\prime }=x_i+\alpha_i\cos R^{-1}t,\qquad v_i^{\prime
}=v_i-\alpha_iR^{-1}\sin
R^{-1}t,&  \label{trbo1} \\
&x_i^{\prime }=x_i+\beta_iR\sin R^{-1}t,\qquad v_i^{\prime
}=v_i+\beta_i\cos R^{-1}t,&  \label{trbo2}
\end{eqnarray}
and  is invariant under  rotations and time translations,
\begin{eqnarray}\label{rotsym}
    &x'_i=x_i\cos \varphi +\epsilon_{ij}x_j\sin\varphi, \qquad
    v'_i=v_i\cos \varphi +\epsilon_{ij}v_j\sin\varphi,&\\
    &t'=t-\gamma.&\label{timetr}
\end{eqnarray}
Newton-Hooke translations (\ref{trbo1}) and boosts (\ref{trbo2}) are
related via a time translation: the former are transformed into the
latter via a shift $t\rightarrow t-\frac{\pi}{2}R$ accompanied with
the change of the transformation parameters $\alpha_i\rightarrow
R\beta_i$.

These symmetry transformations are generated by the vector fields
\begin{equation}
    {\cal X}_{P_i}=\cos
    R^{-1}t\frac{\partial}{\partial x_i}
    -R^{-1}\sin R^{-1}t \frac{\partial}{\partial v_i},\qquad
    {\cal X}_{K_i} =  R \sin R^{-1}t \frac{\partial}{\partial x_i}+\cos
    R^{-1}t\frac{\partial}{\partial v_i},\label{gen1}
\end{equation}
\begin{equation}
    {\cal X}_J=\epsilon_{ij}\left(x_j\frac{\partial}{\partial
    x_i}+v_j\frac{\partial}{\partial v_i}\right),\qquad
     {\cal X}_H = -\frac{\partial}{\partial t}.\label{gen2}
\end{equation}
The algebra of these vector fields is
\begin{equation}\label{alg01}
    [{\cal X}_H, {\cal X}_{K_i}]=-{\cal X}_{P_i},\qquad
    [{\cal X}_H,{\cal X}_{P_i}]=+\frac{1}{R^2}{\cal X}_{K_i},\qquad
    [{\cal X}_{K_i},{\cal X}_{K_j}]=0,\qquad
    [{\cal X}_{P_i},{\cal X}_{P_j}]=0,
\end{equation}
\begin{equation}\label{alg02}
    [{\cal X}_{K_i},{\cal X}_{P_j}]=0,\qquad
    [{\cal X}_J,{\cal X}_{P_i}]=\epsilon_{ij}{\cal X}_{P_j},\qquad
    [{\cal X}_J,{\cal X}_{K_i}]=\epsilon_{ij}{\cal X}_{K_j}.
\end{equation}
There are no central charges in this realization, and it coincides
(up  to the factor $i$) with the algebra of Section 2 but without
central charges.

In correspondence with Eqs. (\ref{trbo1}), (\ref{trbo2}),  in chiral
basis Newton-Hooke translation and boost transformations
 have a form
\begin{equation}\label{NHchiral*}
    X^\pm{}'_i=X^\pm_i+\alpha^\pm_i(t),
\end{equation}
where
\begin{equation}\label{albet+-}
    \alpha^\pm_i(t)=\Delta^\pm_{ij}(\alpha_j \pm R\epsilon_{jk}\beta_k),
\end{equation}
\begin{equation}\label{Del+-}
    \Delta^\pm_{ij}(t)=\delta_{ij}\cos
    \frac{t}{R}\mp\epsilon_{ij}\sin \frac{t}{R}.
\end{equation}
The $\Delta^\pm_{ij}(t)$ are rotation matrices, $\Delta_{ij}(t)\in
SO(2)$,
and the $\alpha^\pm_i(t)$ defined by (\ref{albet+-}) satisfy
equations
\begin{equation}\label{alpha+-t}
    \dot{\alpha}^\pm_i(t)=\mp\frac{1}{R}\epsilon_{ij}\alpha^\pm_j(t).
\end{equation}

In terms of  variables (\ref{lambdai}) the generators of
Newton-Hooke translations and boosts are given by
\begin{equation}\label{PKchiral}
    P_i=\lambda^+_j\Delta^+_{ji}(t)+\lambda^-_j\Delta^-_{ji}(t),\qquad
    K_i=R\left(\lambda^-_j\Delta^-_{jk}(t)-\lambda^+_j\Delta^+_{jk}(t)\right)
    \epsilon_{ki}.
\end{equation}
Due to Eq. (\ref{lamchi}), the set of chiral constraints is
invariant with respect to these transformations,
\begin{equation}\label{PKchi}
    \{P_i,\chi^\pm_j\}=\{K_i,\chi^\pm_j\}=0.
\end{equation}
Angular momentum generating rotation symmetry is
\begin{equation}\label{Jchir}
    J=\epsilon_{ij}(X^+_iP^+_j+X^-_iP^-_j).
\end{equation}
Note a similar structure which have angular momentum (\ref{Jchir})
and total Hamiltonian (\ref{Htotchir}). Relations (\ref{PKchi})
and  $\{J,\chi^\pm_i\}=\epsilon_{ij}\chi^\pm_j$,
$\{H,\chi^\pm_i\}=\pm\frac{1}{R}\epsilon_{ij}\chi^\pm_j$
explicitly show the invariance of the physical subspace given by
the constraints (\ref{chi+}), (\ref{chi-}) under the Newton-Hooke
transformations.

In the reduced phase space described by variables $X^+_i$ and
$X^-_i$ (when $\rho^2\neq 1$) with symplectic structure (\ref{XX++}),
(\ref{XX--}), the angular momentum is
\begin{equation}\label{Jchiral}
    J=\frac{1}{2}\left(\mu_+{X^+_i}^2-\mu_-{X^-_i}^2\right),
\end{equation}
cf. (\ref{H+-}).

The integrals of motion $H$, $J$, ${ P}_i$ and ${ K}_i$ generate the
algebra,
\begin{equation}\label{alg0}
\{H,J\}=0,
\end{equation}
\begin{equation}  \label{alg1}
\{H, K_i\}=-P_i,\qquad \{H,P_i\}=\frac{1}{R^2}K_i,
\end{equation}
\begin{equation}\label{alg2}
    \{J,P_i\}=\epsilon_{ij}P_j,\qquad
    \{J,K_i\}=\epsilon_{ij}K_j,
\end{equation}
\begin{equation}  \label{alg3}
    \{K_i,P_j\}=m\delta_{ij},\qquad
    \{K_i,K_j\}=-\kappa\epsilon_{ij},
\end{equation}
\begin{equation}\label{alg4}
     \{P_i,P_j\}=-\frac{\kappa}{R^2}
    \epsilon_{ij}.
\end{equation}
This is a classical analog of the exotic NH$_3$ algebra
(\ref{NH3ex})--(\ref{PP}) with parameters $m$ and $\kappa$ playing
the role of the central charges $Z$ and $\tilde{Z}$, respectively.
Making use of the explicit form of integrals $H$, $J$, ${ P}_i$
and ${ K}_i$, one finds that on the constraint surface the
Casimirs (\ref{CasNH1}) and (\ref{CasNH2}) take zero values. As a
result, in the non-critical case the Hamiltonian and angular
momentum are represented in terms of NH translations and boosts
generators as
\begin{eqnarray}\label{HKPCas}
    H&=&\frac{1}{m(1-\rho^2)}\left(\frac{1}{2}\left(P_i^2+R^{-2}K_i^2\right)-\rho R^{-1}\epsilon_{ij}K_iP_j\right),\\
    J&=&\frac{1}{m(1-\rho^2)}\left(\epsilon_{ij}K_iP_j-\frac{1}{2}\rho R\left(P_i^2+R^{-2}K_i^2\right)\right).\label{JKPCas}
\end{eqnarray}

 The linear combinations of $P_i$ and $K_i$ corresponding to chiral
 generators (\ref{NHchNch}) are given here in terms of variables $\lambda^\pm_i$,
 \begin{equation}\label{calJlam}
    {\cal J}^\pm_i=\mp R\lambda^\pm_j\Delta^\pm_{ji}(t).
\end{equation}
The equations
\begin{equation}\label{partJ}
    \frac{\partial {\cal J}^\pm_i}{\partial
    t}=\pm\frac{1}{R}\epsilon_{ij}{\cal J}^\pm_j
\end{equation}
guarantee that they are integrals of motion, $\frac{d}{dt}{\cal
J}^\pm_i=\frac{\partial {\cal J}^\pm_i}{\partial t}+\{{\cal
J}^\pm_i,H\}=0$. Note also that on the constraint surface
(\ref{chi+}), (\ref{chi-}) these integrals can be presented in the
form
\begin{equation}\label{JXpm}
    {\cal J}^\pm_i=R\mu_\pm\epsilon_{ij}X_j^\pm (0).
\end{equation}
Together with ${\cal J}^\pm=\frac{1}{2}(J\pm RH)$, where $H$
corresponds to the total Hamiltonian, integrals (\ref{JXpm})
generate classical analog of the exotic NH$_3$ algebra
(\ref{NHchiral}) in the chiral form, in which
\begin{equation}\label{ZZmu}
    Z^+=-R^2\mu_+,\qquad
    Z^-=R^2\mu_-.
\end{equation}

In the critical case corresponding to $\rho=1$ ($\mu_-=0$) or
$\rho=-1$ ($\mu_+=0$), the symmetry of the system is reduced to the
centrally extended NH$_2$ algebra generated by ${\cal J}^+$, ${\cal
J}^+_i$ and $Z^+$, or by ${\cal J}^-$, ${\cal J}^-_i$ and $Z^-$.
Here the Hamiltonian and angular momentum on the one hand, and
translations and boosts generators on the other hand are linearly
dependent,
\begin{eqnarray}\label{CritHJKP}
    \mu_-=0&:&\quad H=R^{-1}J = \frac{1}{2m}P_i^2,\qquad P_i=-R^{-1}\epsilon_{ij}K_j,\\
    \mu_+=0&:&\quad H=-R^{-1}J = \frac{1}{2m}P_i^2,\qquad
    P_i=R^{-1}\epsilon_{ij}K_j.\label{CritHJKP1}
\end{eqnarray}
Note that the sign of $H$ is defined by the sign of the mass
parameter $m$.

In the non-chiral basis the generators of Newton-Hooke translations
and boosts are given in terms of observable variables
(\ref{obserNC}),
\begin{equation}\label{PKNCobs}
    P_i={\cal P}_i\cos\frac{t}{R}+\frac{1}{R}{\cal K}_i\sin\frac{t}{R},\qquad
    K_i={\cal K}_i\cos\frac{t}{R}-R{\cal P}_i\sin\frac{t}{R},
\end{equation}
while the angular momentum is
\begin{equation}\label{Jncir}
    J=\epsilon_{ij}(x_ip_j+v_i\pi_j).
\end{equation}
Together with total Hamiltonian (\ref{HtotNC}) they generate
classical algebra (\ref{alg0})--(\ref{alg4}).

In  the reduced phase space with coordinates $x_i,$ $p_i$,
integrals $P_i$, $K_i$ and $J$  take the form
    \begin{equation}  \label{Pi}
    P_i={\cal P}_i\cos\frac{t}{R}
    +\frac{1}{R}\tilde{\cal K}_i\sin\frac{t}{R},\qquad
    K_i= \tilde{\cal K}_i\cos\frac{t}{R}
    -R{\cal P}_i\sin\frac{t}{R},
\end{equation}
\begin{equation}\label{Jred}
    J=\frac{m^2}{2\kappa}\left(\left(x_i+\frac{\kappa}{m^2}\epsilon_{ij}p_j\right)^2
    -(1-\rho^2)x_i^2\right),
\end{equation}
where
\begin{equation}\label{tilKi}
     \tilde{\cal K}_i=mx_i\left(1-\frac{1}{2}\rho^2\right)+
    \frac{\kappa}{m}\epsilon_{ij}p_j.
\end{equation}
Together with Hamiltonian (\ref{Hredxp}) they satisfy the same
classical algebra (\ref{alg0})--(\ref{alg4}) with respect to the
reduced symplectic structure (\ref{xxpx}).

\subsection{SO(3) and SO(2,1) symmetry}\label{addsymduality0}

Let us now show that our system has an additional symmetry. For
the chiral Lagrangian this additional symmetry is manifest, its
explicit form  depends of the phase we consider.

Define the dimensionless (rescaled) coordinates
\begin{equation}\label{YYY}
\sqrt{|\mu^+|}X^+_1=Y_1,\quad \sqrt{|\mu^+|}X^+_2=Y_2,\quad
\sqrt{|\mu^-|}X^-_1=Y_3,\quad \sqrt{|\mu^-|}X^-_2=Y_4.
\end{equation}
In terms of these coordinates the chiral Lagrangian is written as
\begin{equation}\label{Ly}
    {\cal L}_{ch}=-\frac{1}{2}
\left(\dot{Y}^T\W Y+\frac{1}{R}{Y}^T\h Y\right)=-\frac{1}{2}
\left(\dot{Y}^A\W_{AB} Y^B+\frac{1}{R}{Y}^A\h_{AB} Y^B\right),
\end{equation}
where
\begin{equation}\label{Ly*}
    \eta_{AB}=
\left(
\begin{array}{cccc}
  \varepsilon_+& . & . & . \\
  . & \varepsilon_+ & . & . \\
  . & . & \varepsilon_- & . \\
  . & . & . & \varepsilon_- \\
\end{array}%
\right),\qquad \W_{AB}=
\left(%
\begin{array}{cccc}
  .& \varepsilon_+ & . & . \\
  -\varepsilon_+ & . & . & . \\
  . & . & . & -\varepsilon_- \\
  . & . & \varepsilon_- & . \\
\end{array}%
\right),
\end{equation}
$\varepsilon_\pm$ are the signs of $\mu_\pm$, and the matrices
$\eta$ and $\W$ satisfy the relations $\h^2=1,$ $\W^T=-\W,$
 $\W\W=-1$.
The potential term is invariant under transformations
$Y\rightarrow OY$, $O^T\eta O=\eta$, which are the SO(4) rotations
in the subcritical case with $\varepsilon_+\varepsilon_-=+1$, and
are the pseudo-rotations SO(2,2)$\sim$AdS$_3$ in supercritical
sector characterized by the relation
$\varepsilon_+\varepsilon_-=-1$. On the other hand, the first,
kinetic term is invariant under the Sp(4) transformations with
symplectic metric $\W$, $Y\rightarrow CY$, $C^T\W C=\W$.  The
symmetry of Lagrangian corresponds to the intersection of SO(4)
(or, SO(2,2)) and Sp(4). Considering an infinitesimal
transformation $\delta Y=\omega Y$, we find that the constant
matrix $\omega$ has to  satisfy equations
$\omega^T\eta+\eta\omega=0$, $\omega^T\Omega+\Omega\omega=0$. A
simple analysis of these equations with subsequent application of
the Noether theorem  results finally in the set of the four
integrals of motion which are $H$, $J$ and
\begin{equation}
I_1=\frac 12\varepsilon _{+}\sqrt{\left\vert \frac{\mu _{-}}{\mu
_{+}}\right\vert
}\left( P_{2}^{+}X_{2}^{-}-P_{1}^{+}X_{1}^{-}\right) +\frac 12\varepsilon _{-}\sqrt{%
\left\vert \frac{\mu _{+}}{\mu _{-}}\right\vert }\left(
X_{1}^{+}P_{1}^{-}-X_{2}^{+}P_{2}^{-}\right) ,  \label{a1}
\end{equation}
\begin{equation}
I_2=-\frac 12\varepsilon _{+}\sqrt{\left\vert \frac{\mu _{-}}{\mu _{+}}%
\right\vert }\left( P_{2}^{+}X_{1}^{-}+P_{1}^{+}X_{2}^{-}\right)
+\frac 12\varepsilon _{-}\sqrt{\left\vert \frac{\mu _{+}}{\mu _{-}}\right\vert }%
\left( X_{1}^{+}P_{2}^{-}+X_{2}^{+}P_{1}^{-}\right).  \label{a2}
\end{equation}
On the surface of the constraints (\ref{chi+}), (\ref{chi-}) these
two new integrals  take a form
\begin{equation}\label{II12}
    I_1=\frac 12\sqrt{|\mu_+\mu_-|}\,(X^+_1X^-_2+X^+_2X^-_1),\qquad
    I_2=\frac 12\sqrt{|\mu_+\mu_-|}\,(X^+_2X^-_2-X^+_1X^-_1).
\end{equation}
In subcritical case integral $I_1$ generates 2D rotations in the
planes $(Y_1,Y_3)$ and $(Y_2,Y_4)$, while $I_2$ generates rotations
in the planes $(Y_1,Y_4)$ and $(Y_2,Y_3)$. In supercritical case the
2D rotations are changed for 2D Lorentz transformations in the same
planes. Being time-independent, these integrals commute with the
Hamiltonian, $\{H,I_1\}=\{H,I_2\}=0$, and together with rescaled
angular momentum
\begin{equation}\label{I3J}
    I_3=\frac{1}{2}J
\end{equation}
they generate the Lie algebra
\begin{equation}\label{so3so21}
    \{I_1,I_2\}=\varepsilon_+\varepsilon_- I_3,\qquad
    \{I_3,I_1\}=I_2,\qquad
    \{I_2,I_3\}=I_1.
\end{equation}
In subcritical case (\ref{so3so21}) is identified as a classical
rotation symmetry algebra $so(3)$, while in supercritical case it
is a Lorentz algebra $so(2,1)$. Note also that the brackets of the
integrals $I_{1,2}$ with the Newton-Hooke translation and boost
generators are reduced to some linear combinations of the latter.

\section{Reduced phase space quantization}\label{reducedquant}

In correspondence with classical relations (\ref{XX++}),
(\ref{XX--}), in subcritical case $\rho^2<1$ with $\mu_+>0$,
$\mu_->0$, one defines two sets of creation-annihilation oscillator
operators
\begin{equation}
    a_{+}=\sqrt{\frac{|\mu_+|}{2}}\, (X^+_2+ i X^+_1),\qquad
    a_-=\sqrt{\frac{|\mu_-|}{2}}\, (X^-_1+ i X^-_2),\qquad
    a_\pm^\dagger=\left(a_\pm\right)^\dagger,
 \label{a+a-}
\end{equation}
\begin{equation}\label{aaaa}
[a_+,a^\dagger_+]=1,\qquad [a_-,a^\dagger_-]=1,\qquad
[a_+,a_-]=[a_+,a^\dagger_-]=0.
\end{equation}
We put here  parameters $\mu_\pm$ under the modulus sign having in
mind further generalization. Let us construct Hamiltonian and
angular momentum operators fixing the symmetrized ordering in
(\ref{H+-*}) and (\ref{Jchiral}). We get
\begin{equation}\label{HX+X-redsub}
    H=R^{-1}\left(a^\dagger_+a_++
    a^\dagger_-a_-+1\right),\qquad
    J=a^\dagger_+a_+ -
    a^\dagger_-a_-\,.
\end{equation}
Quantum system is an ordinary planar isotropic oscillator of
frequency $R^{-1}$. Quantum Newton-Hooke chiral operators ${\cal
J}_i^\pm$  are constructed via (\ref{JXpm}), (\ref{a+a-}) and
relations inverse to (\ref{X+-motion}),
\begin{equation}\label{JJJJ1}
    \mathcal{J}_{1}^{+}=R\sqrt{\frac{\mu _{+}}{2}}\left(e^{ -itR^{-1}}
    a^\dagger_+ +e^{ itR^{-1}} a_+\right),\quad
     \mathcal{J}_{2}^{+}=-iR\sqrt{\frac{\mu
    _{+}}{2}}\left( e^{ -itR^{-1}}
     a^\dagger_+ - e^{ itR^{-1}} a_+
    \right),
\end{equation}
\begin{equation}\label{JJJJ2}
    \mathcal{J}_{1}^{-}=iR\sqrt{\frac{\mu _{-}}{2}}\left( e^{ -itR^{-1}}
     a^\dagger_--e^{ itR^{-1}} a_-
    \right),\quad
     \mathcal{J}_{2}^{-}=-R\sqrt{\frac{\mu
    _{-}}{2}}\left(e^{ -itR^{-1}}a^\dagger_- + e^{ itR^{-1}} a_-\right).
\end{equation}
They form the exotic NH$_3$ algebra from Section 2. Energy takes
positive values, $E_{n_+,n_-}=R^{-1}(n_++n_-+1)$, while angular
momentum can take values of both signs, $j_{n_+,n_-}=n_+-n_-$, where
$n_\pm=0,1,\ldots$ are the eigenvalues of the number operators
$N_\pm=a_\pm^\dagger a_\pm$, $N_\pm|n_+,n_-\rangle=n_\pm
|n_+,n_-\rangle$.

In subcritical phase $\rho^2<1$ with $\mu_+<0$, $\mu_-<0$, the
operators $a_+$ and $a_-$ are realized as in (\ref{a+a-}) with the
substitution $X^+_1\leftrightarrow X^+_2$, $X^-_1\leftrightarrow
X^-_2$. This provokes the change of global signs in Hamiltonian and
angular momentum,
\begin{equation}\label{HX+X-redsub*}
    H=-R^{-1}\left(a^\dagger_+a_+ +
    a^\dagger_-a_-+1\right),\qquad
    J=-a^\dagger_+a_+ + a^\dagger_-a_-\,.
\end{equation}

In the supercritical case $\rho>1$ ($\mu_+>0$, $\mu_-<0$), operator
$a_+$ is defined as in (\ref{a+a-}) while $ a_-$ is obtained via the
substitution $X^-_1\leftrightarrow X^-_2$. Here
\begin{equation}\label{Hrho>1}
    H=R^{-1}\left(a^\dagger_+a_+ -
    a^\dagger_-a_-\right),\qquad
    J=a^\dagger_+a_+ + a^\dagger_- a_- +1.
\end{equation}
This is an exotic oscillator with interchanged Hamiltonian and
angular momentum. Energy can take positive and negative values,
$E_{n_+,n_-}=R^{-1}(n_+-n_-)$, $n_\pm=0,1,\ldots$, and is not
restricted from below. Angular momentum can take only positive
values, $j_{n_+,n_-}=n_++n_-+1$, i.e. the both modes are `right'.

In the supercritical case $\rho<-1$ ($\mu_+<0$, $\mu_->0$), the role
of the modes is changed: the energy for the mode $X^+$ is negative,
while for $X^-$ is positive. In this case $a_-$ is defined as in
(\ref{a+a-}), while $ a_+$ is realized via the change
$X^+_1\leftrightarrow X^+_2$. As a result,
\begin{equation}\label{Hrho<-1}
    H=R^{-1}\left(-a^\dagger_+ a_+ +
    a^\dagger_-a_-\right),\qquad
    J=-\left(a^\dagger_+a_+ +
    a^\dagger_-a_-+1\right).
\end{equation}
We have, again, an exotic oscillator, but now with both modes to be
`left'.

The expressions for Hamiltonian and angular momentum for all  four
cases can be unified as
\begin{equation}\label{HJuni}
    H=R^{-1}\left(\varepsilon_+a^\dagger_+a_+ +
    \varepsilon_-a^\dagger_- a_-+\frac{1}{2}(\varepsilon_++\varepsilon_-)\right),\qquad
    J=\varepsilon_+a^\dagger_+ a_+ -
    \varepsilon_-a^\dagger_-a_-+\frac{1}{2}(\varepsilon_+-\varepsilon_-),
\end{equation}
where $\varepsilon _{\pm }=\mathrm{sign}\,\mu _{\pm }$.

The boost and translation generators are given in terms of the
chiral integrals,
\begin{equation}\label{RedQ1}
    K_{i}=-\epsilon _{ij}\left(
    \mathcal{J}_{j}^{+}+\mathcal{J}_{j}^{-}\right),
    \qquad P_{i}=-\frac{1}{R}\left( \mathcal{J}_{i}^{+}-\mathcal{J}%
    _{i}^{-}\right),
\end{equation}
which can be presented in a compact form  generalizing
(\ref{JJJJ1}), (\ref{JJJJ2}) for the case of arbitrary  signs of
$\mu_\pm$,
\begin{equation}\label{RedQ2}
    \mathcal{J}_{j}^{+}=\left( -i\right) ^{j-1}\left( \varepsilon
    _{+}\right) ^{j+1/2}R\sqrt{\frac{|\mu _{+}|}{2}}\left(
     e^{ -i\varepsilon_+tR^{-1}}a^\dagger_+
    -\left( -1\right) ^j\varepsilon
    _{+}e^{i\varepsilon_+tR^{-1}}a_+\right),
\end{equation}
\begin{equation}\label{RedQ3}
    \mathcal{J}_{j}^{-}=\left( i\right) ^{j}\left( \varepsilon
    _{-}\right) ^{j-1/2}R\sqrt{\frac{|\mu _{-}|}{2}}\left(
     e^{ -i\varepsilon_-tR^{-1}}a^\dagger_-
     +\left( -1\right) ^j\varepsilon
    _{-}e^{i\varepsilon_-tR^{-1}}a_- \right),
\end{equation}
where we imply $(-1)^{1/2}=i$.

The ladder operators $I_+=I_1+i I_2$ and  $I_-=I_+^\dagger$ of the
additional SO(3) ($\varepsilon_+\varepsilon_-=+1$), or SO(2,1)
($\varepsilon_+\varepsilon_-=-1$) symmetry,
\begin{equation}\label{I+I+I3}
    [I_+,I_-]=\varepsilon_+\varepsilon_-i I_3,\qquad
    [I_3,I_\pm]=\pm iI_\pm,
\end{equation}
 corresponding to
complex combinations of the classical integrals (\ref{II12}) are
given by
\begin{equation}\label{HJunia1}
\varepsilon_+\varepsilon_-=+1:\qquad
I_1+i\varepsilon_+I_2=a^\dagger_+a_-,
\end{equation}
\begin{equation}\label{HJunia2}
\varepsilon_+\varepsilon_-=-1:\qquad
I_1+i\varepsilon_+I_2=ia^\dagger_+a_-^\dagger.
\end{equation}

In the critical phase with $\mu_-=0$ we have a one-mode oscillator
with
   $\quad E_{n_+}=\varepsilon_+
    R^{-1}\left(n_++\frac{1}{2}\right),$
    $j_{n_+}=\varepsilon_+\left(n_++\frac{1}{2}\right)$, while for
    $\mu_+=0$ we have
    $ E_{n_-}=\varepsilon_-R^{-1}\left(n_-+\frac{1}{2}\right),$
    $j_{n_-}=-\varepsilon_-\left(n_-+\frac{1}{2}\right)$.
The generators of the corresponding centrally extended NH$_2$
symmetry are given, respectively, by  Eq. (\ref{RedQ2}) or
(\ref{RedQ3}), and by Eq. (\ref{HJuni}) if we put in the latter the
parameter $\varepsilon_-$ or $\varepsilon_+$ equal to zero.

\section{Duality}\label{addsymduality}

Let us clarify the relation between the sub- and super-critical
phases in the light of duality transformation, which can be
presented in three equivalent ways,
\begin{equation}\label{duality3}
    \rho\rightarrow \rho^{-1};\qquad  mR\leftrightarrow
    \kappa;\qquad \mu_\pm\rightarrow \pm\mu_\pm.
\end{equation}
The duality induces a mutual transformation between the sub- and
super-critical phases, and between two critical phases given by
$\mu_+=0$, $\mu_-<0$ and $\mu_+=0$, $\mu_->0$, while it leaves
invariant the critical phases with $\mu_-=0$, see Figure 1.
\begin{figure}[h]
\begin{center}
  \includegraphics[width=.6\textwidth]{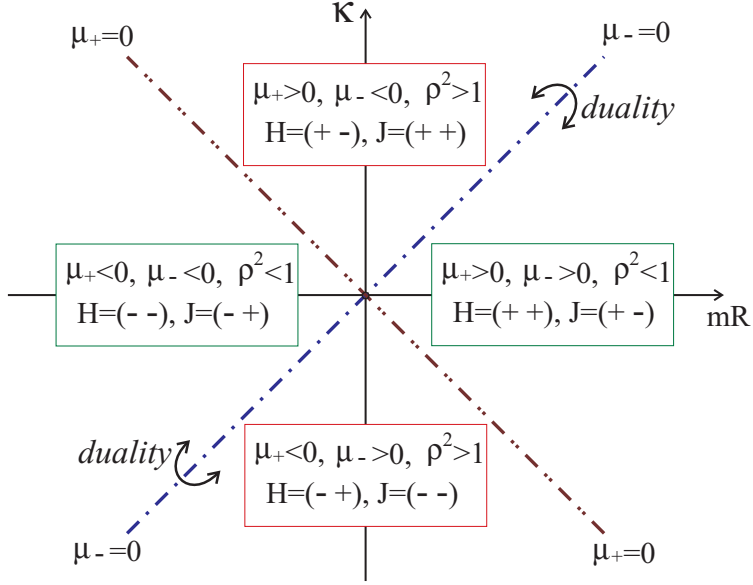}\\
  \caption{Phases and duality.}\label{duality_fig}
     \end{center}
\end{figure}
In accordance with Eqs. (\ref{HJuni})--(\ref{RedQ3}), duality
transformation (\ref{duality3}) induces the transformations
\begin{equation}\label{JHdual}
    J\leftrightarrow RH,
\end{equation}
\begin{eqnarray}\label{PBo1}
    P_i&\rightarrow& P_i'=\frac{1}{2}\left(
    (P_1+P_2)+(-1)^iR^{-1}(K_1+K_2)\right),\\
    K_i&\rightarrow& K_i'=\frac{1}{2}\left((-1)^{i+1}
    (K_1-K_2)+R(P_2-P_1)\right).\label{PBo4}
\end{eqnarray}
In terms of chiral integrals (\ref{NHchNch}), these
transformations are equivalent to
\begin{equation}\label{chirJJJ}
    {\cal J}^\pm\rightarrow \pm{\cal J}^\pm,\qquad
    {\cal J}^+_i\rightarrow {\cal J}^+_i,\qquad
    {\cal J}^-_1\rightarrow {\cal J}_2^-,\qquad
     {\cal J}^-_2\rightarrow {\cal J}_1^-.
\end{equation}
Having in mind also (\ref{ZZmu}), we have
\begin{equation}\label{ZZZZ}
    Z^\pm\rightarrow \pm Z^\pm.
\end{equation}
Using relations (\ref{chirJJJ}), (\ref{ZZZZ}), we conclude that
the duality does not change the algebra (\ref{NHchiral}) of
integrals of motion, and so, it is an automorphism of the exotic
Newton-Hooke algebra.

On the other hand, the sub- and super-critical phases are
essentially different from the viewpoint of the structure of the
energy and angular momentum levels. In correspondence with
(\ref{HJuni}), the energy levels and angular momentum values are
\begin{equation}\label{Esub+-}
    E_{n_+,n_-}=\frac{1}{R}\left(\varepsilon_+n_+
    +\varepsilon_-n_-+\frac{1}{2}(\varepsilon_++\varepsilon_-)\right),\qquad
    j_{n_+,n_-}=\varepsilon_+n_+-\varepsilon_-n_-+
    \frac{1}{2}(\varepsilon_+-\varepsilon_-),
\end{equation}
where $n_\pm=0,1,\ldots$. Every such a level in subcritical case
$\varepsilon_+\varepsilon_-=+1$ has a finite degeneration equal to
$2j_{{}_<}+1$, where
\begin{equation}\label{j<def}
    j_{{}_<}=\frac{1}{2}\left(n_++n_-\right),
\end{equation}
and corresponding energy eigenstates are characterized by the
quantum numbers $n_+$ and $n_-$ laying on the straight lines
restricted by vertical and horizontal axes, see Figure 2. The
$so(3)$ ladder operators given by Eq. (\ref{HJunia1}) act along
these lines. On the line with $n_++n_-=2j_{{}_<}$, the $so(3)$
Casimir operator $C_{so(3)}=I_1^2+I_2^2+I_3^2$ takes a value
$j_{{}_<}(j_{{}_<}+1)$.

\begin{figure}[h]
\begin{center}
  \includegraphics[width=.55\textwidth]{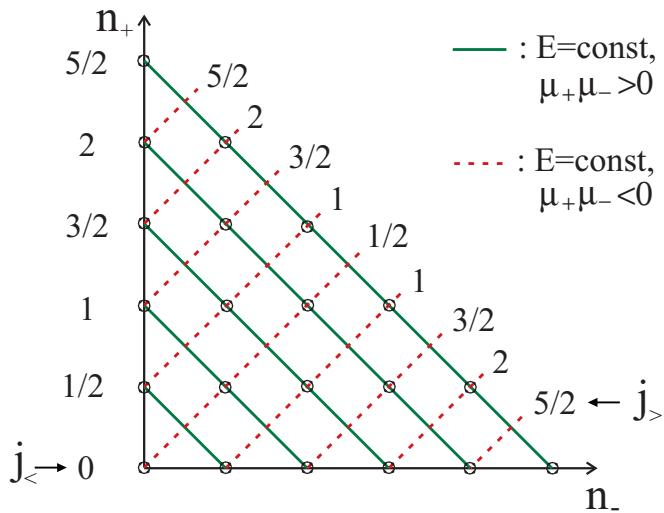}\\
  \caption{Constant energy levels.}\label{graf_fig}
   \end{center}
\end{figure}

In supercritical phase  $\varepsilon_+\varepsilon_-=-1$, and
constant energy levels correspond to the dotted straight lines
$n_+-n_-=const$ on Figure 2.  The $so(2,1)$ ladder operators
(\ref{HJunia2}) act along them. Every energy level is infinitely
degenerated, and the $so(2,1)$ Casimir operator
$C_{so(2,1)}=I_1^2+I_2^2-I_3^2$ takes a value
$-j_{{}_>}(j_{{}_>}-1)$, where
\begin{equation}\label{j>so21}
    j_{{}_>}=\frac{1}{2}\left(|n_+-n_-|+1\right).
\end{equation}
Every such a straight line corresponds to the infinite-dimensional
half-bounded unitary representation of $so(2,1)$ algebra, on which
the $so(2,1)$ compact generator $I_3$ takes the values
$i_3=\varepsilon_+\frac{1}{2}(n_++n_-+1)$. This can be presented
equivalently as  $i_3=\varepsilon_+ (j_{{}_>}+k),$ where
$k=n_-=0,1,\ldots$ for $n_+\geq n_-$, and $k=n_+=0,1,\ldots$ for
$n_-\geq n_+$ . Therefore, we have here the so called discrete
series of representations  $D^+_{j_{{}_>}}$ (for $\varepsilon_+=+1$)
and $D^-_{j_{{}_>}}$ (for $\varepsilon_+=-1$) of $SL(2,R)$ in
terminology of Bargmann \cite{BarSL2R}.

 We
conclude that the duality transformation (\ref{duality3}) provokes a
mutual change of compact, $so(3)$, and noncompact, $so(2,1)$,
symmetries in accordance with Eq. (\ref{I+I+I3}). This is
accompanied by a radical change of the energy degeneration from the
finite to the infinite, or inversely.

\section{Wave equations}\label{waveEquat}

Here we shall quantize the system using the method of
Gupta-Bleuler, that will allow us to construct the set of wave
equations realizing the exotic Newton-Hooke symmetry. Using them,
in the next Section we shall identify corresponding projective
phases associated with Newton-Hooke translations and boosts
transformations.

\subsection{Chiral basis}

Let us choose two complex linear combinations of the constraints
(\ref{chi+}), (\ref{chi-}) such that

\begin{itemize}

\item they would  commute between themselves,

\item
  at the quantum level they would have
a nature of annihilation operators having nontrivial kernels,
\item
 their set would be consistent with dynamics,
i.e. their commutators with Hamiltonian would be proportional to
the chosen combinations of the constraints.

\end{itemize}

\noindent
 Then at the quantum level these linear combinations will
separate a physical subspace of the system.

 In correspondence with the structure of the brackets of the
chiral constraints (\ref{chi+}), (\ref{chi-}), for $\rho^2\neq 1$ it
is necessary to distinguish four cases in dependence  on the signs
of $\mu_\pm$. For $\mu_+>0$, $\mu_->0$, the suitable choice of the
linear combinations of the constraints is
\begin{equation}\label{chir+-sub}
\chi^+_1+i\chi^+_2\approx 0,\qquad \chi^-_1-i\chi^-_2\approx 0.
\end{equation}
 In
coordinate representation they transform into quantum equations
\begin{equation}\label{++case}
    \left(\partial_{\bar{\cal Z}^+}+\frac{\mu_+}{4}{\cal Z}^+\right)\Psi
    =0,\qquad
    \left(\partial_{{\cal Z}^-}
    +\frac{\mu_-}{4}\bar{\cal Z}^-\right)\Psi
    =0,
\end{equation}
where $\partial_{\bar{\cal Z}^+}=\partial/\partial \bar{\cal Z}^+$,
$\partial_{{\cal Z}^-}=
\partial/\partial {\cal
    Z}^-$, and
 we have introduced two independent complex
chiral variables
\begin{equation}\label{Z+Z-}
    {\cal Z}^+=X^+_1+iX^+_2,\qquad
    {\cal Z}^-=X^-_1+iX^-_2.
\end{equation}
The dynamics is given by the Schr\"odinger equation
\begin{equation}\label{Schchir}
    \left(i\partial_t -\frac{1}{R}\left(
    {\cal Z}^+\partial_{{\cal Z}^+}
    -\bar{\cal Z}^+\partial_{\bar{\cal Z}^+}
    -{\cal Z}^-\partial_{{\cal Z}^-}
    +\bar{\cal Z}^-\partial_{\bar{\cal Z}^-}
    \right)\right)\Psi=0,
\end{equation}
where the linear differential operator in complex chiral variables
is a quantum analog of the total chiral Hamiltonian
(\ref{Htotchir}).  Eqs. (\ref{++case}) and (\ref{Schchir}) represent
the set of wave equations for our particle system with exotic
Newton-Hooke symmetry.

The solution of the quantum equations (\ref{++case}) is
\begin{equation}\label{psiphys}
    \Psi_{phys}=\exp\left(
    -\frac{\mu_+}{4}|{\cal Z}^+|^2
    -\frac{\mu_-}{4}|{\cal Z}^-|^2
    \right)\psi({\cal Z}^+,\bar{\cal Z}^-).
\end{equation}
The action of the Hamiltonian and angular momentum operators on the
states (\ref{psiphys}) is reduced to
\begin{equation}\label{HJPsi}
    H\Psi_{phys}=\exp(.)\frac{1}{R}\left(
    {\cal Z}^+\partial_{{\cal Z}^+}
    +\bar{\cal Z}^-\partial_{\bar{\cal Z}^-}\right)\psi({\cal Z}^+,\bar{\cal
    Z}^-),
\end{equation}
\begin{equation}\label{HJPsi*}
    J\Psi_{phys}=\exp(.)\frac{1}{R}\left(
    {\cal Z}^+\partial_{{\cal Z}^+}
    -\bar{\cal Z}^-\partial_{\bar{\cal Z}^-}\right)\psi({\cal Z}^+,\bar{\cal
    Z}^-).
\end{equation}
The energy and angular momentum eigenfunctions are given by the
states (\ref{psiphys}) with monomial wave functions $\psi({\cal
Z}^+,\bar{\cal Z}^-)$,
\begin{equation}\label{psinn}
    \psi_{n_+,n_-}({\cal
    Z}^+,\bar{\cal Z}^-)=
    \frac{1}{\sqrt{n_+!n_-!}}\left(\sqrt{\frac{\mu_+}{2}}{\cal Z}^+\right)^{n_+}
    \left(\sqrt{\frac{\mu_-}{2}}\bar{\cal
    Z}^-\right)^{n_-},
    \qquad n_+,n_-=0,1,\ldots,
\end{equation}
which correspond to the energy and angular momentum eigenvalues
$E_{n_+,n_-}=\frac{1}{R}(n_++n_-)$, $j_{n_+,n_-}=n_+-n_-$. Note
that a simple  quantum shift in energy [equal to
$-\frac{1}{R}\hbar$, $\hbar=1$] in comparison with the reduced
phase space quantization scheme is related to our choice of the
normal quantum ordering: here we take a Hamiltonian operator
corresponding to the classical expression (\ref{Htotchir}).

 These
results are in a complete correspondence with reduced phase space
quantization scheme [up to inessential shift in energy], and what we
get here is the holomorphic representation for the 2D oscillator
system.

The quantization scheme is generalized directly for other three
cases of the signs $\varepsilon_+$ and $\varepsilon_-$ of the
parameters $\mu_+$ and $\mu_-$. The wave equations separating
physical states are given by the linear combinations of the
constraints $\chi^+_i$ and $\chi^-_i$,
\begin{equation}\label{quantconst}
    \left(\chi^+_1+i\varepsilon_+\chi^+_2\right)\Psi=0,\qquad
    \left(\chi^-_1-i\varepsilon_-\chi^-_2\right)\Psi=0.
\end{equation}
In the table we summarize all the cases. \vskip 0.8cm
\begin{tabular}{|c|c|c|c|c|}
  \hline
  Phase & $\varepsilon_+,$ $\varepsilon_- $ & $m,$ $\kappa$ & Constraints & $\psi(.,.)$  \\
  \hline
  $\rho^2<1$& $+1,+1$ & $m>0,$ $|\kappa|<mR$ & $\chi^+_1+i\chi^+_2,$ $\chi^-_1 -i\chi^-_2$& $\overset{}{\psi({\cal Z}^+,\bar{\cal Z}^-)}$ \\
  $\rho^2<1$ & $-1,-1$ & $m<0$, $|\kappa|<mR$ &$\chi^+_1-i\chi^+_2,$ $\chi^-_1 +i\chi^-_2$&  $\psi(\bar{\cal Z}^+,{\cal Z}^-)$ \\
  $\rho^2>1$ & $+1,-1$ & $-\infty<m<\infty$, $\kappa>|m|R$ & $\chi^+_1+i\chi^+_2,$ $\chi^-_1 +i\chi^-_2$& $\psi({\cal Z}^+,{\cal Z}^-)$ \\
  $\rho^2>1$ & $-1,+1$  & $-\infty<m<\infty$, $\kappa<|m|R$ & $\chi^+_1-i\chi^+_2,$
  $\chi^-_1 - i\chi^-_2$& $\psi(\bar{\cal Z}^+,\bar{\cal Z}^-)$ \\
  \hline
\end{tabular}
\vskip 0.8cm \noindent Corresponding solutions describing physical
states have a form similar to (\ref{psiphys}),
\begin{equation}\label{psiphys*}
    \Psi_{phys}=\exp\left(
    -\frac{|\mu_+|}{4}|{\cal Z}^+|^2
    -\frac{|\mu_-|}{4}|{\cal Z}^-|^2
    \right)\psi(.,.),
\end{equation}
with the arguments of the function $\psi$ specified in the table.
The action of $H$ and $J$ on these states has a form similar to
(\ref{HJPsi}), (\ref{HJPsi*}): they are reduced to the sums of two
first order differential operators in variables corresponding to the
arguments of the function $\psi(.,.)$.  The signs appearing before
the operators are indicated on Figure 1, and they correspond to the
signs in Eq. (\ref{HJuni}). In analogs of functions  (\ref{psinn}),
parameters $\mu_\pm$ are changed for their absolute values. The
scalar product is defined in all the cases with a measure
$$
{\cal D}=\frac{|\mu_+\mu_-|}{(2\pi)^2}dX^+_1dX^+_2dX^-_1dX^-_2.
$$
With
respect to such a scalar product wave functions of the form
(\ref{psiphys}), (\ref{psinn}) represent an orthonormal basis in the
physical subspace of the system.

In the critical phase  we have $\mu_+=0$, or $\mu_-$=0, and in
accordance with classical picture one of the quantum complex
equations specifying physical states is changed for
\begin{equation}\label{dX+-}
    \frac{\partial}{\partial X^+_i}\Psi=0,\quad {\rm or}\quad
    \frac{\partial}{\partial X^-_i}\Psi=0.
\end{equation}
The corresponding mode $X^+$ or $X^-$ completely disappears from the
theory, and we get a one mode oscillator in holomorphic
representation.

Note that the quantum equations (\ref{quantconst}) specifying here
the states of the physical subspace have a sense of complex
polarizations for the 2D oscillator treated within the framework
of geometric quantization method \cite{GQ}, while  equations in
(\ref{dX+-}) have a nature of real polarizations.

\subsection{Wave equations in non-chiral basis and flat limit}

Wave equations and physical states in the non-chiral basis can be
directly obtained from those in the chiral basis with taking into
account a phase (unitary) transformation associated with total time
derivative shift (\ref{Nc-Ch}). For example, in supercritical phase
characterized by $\varepsilon_+\varepsilon_-=-1$, physical wave
functions have the form
\begin{equation}\label{physnonCh}
    \Psi_{phys}=\exp\left(-\frac{|\kappa|}{4}\left(v_i^2+\frac{x_i^2}{R^2}\right)
    -\frac{m}{2}\epsilon_{ij}x_iv_j+i\frac{m}{2}x_iv_i\right)\psi(x_1+i\varepsilon_+x_2,v_1+i\varepsilon_+v_2,t).
\end{equation}
They satisfy the constraint wave equations
\begin{equation}\label{V+-Pi+-}
    (V_1+i\varepsilon_+V_2)\Psi(x,v,t)=0,\qquad
    (\Pi_1+i\varepsilon_+\Pi_2)\Psi(x,v,t)=0,
\end{equation}
where $V_i$ and $\Pi_i$ are given by (\ref{constNon1}),
(\ref{constNon2}), and we assume here that
 $p_i=-i\partial/\partial x_i$ and $\pi_i=-i\partial/\partial v_i$.
 The dynamics is given by the Schr\"odinger equation
\begin{equation}\label{HtotNC*}
    \left(i\frac{\partial}{\partial t}
    -iv_j\frac{\partial}{\partial
    x_j}+i\frac{1}{R^2}x_j\frac{\partial}{\partial v_j}
    -\frac{m}{2}\left(
    \frac{x_j^2}{R^2}-v_j^2\right)\right)\Psi=0,
\end{equation}
in which, in accordance with consistency relations
(\ref{NCconsistency}), total Hamiltonian (\ref{HtotNC}) plays  a
role of a quantum Hamiltonian.

With taking into account relations (\ref{chi-nochi}), there is the
following  correspondence between non-chiral, (\ref{constNon1}),
(\ref{constNon2}), and chiral, (\ref{chi+}), (\ref{chi-}),
constraints,
\begin{equation}\label{nonChRel}
    V_i\leftrightarrow (\chi^+_i+\chi^-_i),\qquad
    \Pi_i\leftrightarrow R\epsilon_{ij}(\chi_j^--\chi^+_j).
\end{equation}
This shows that equations (\ref{V+-Pi+-}) are linear combinations of
the quantum chiral constraints (\ref{quantconst}) with
$\varepsilon_+\varepsilon_-=-1$.

In a similar way, one can write down wave equations and physical
states in non-chiral basis for sub-critical phase where
$\varepsilon_+\varepsilon_-=1$. However, from the viewpoint of the
flat limit $R\rightarrow \infty$ (which can be taken in a
subcritical phase only), quantum constraint equations
(\ref{quantconst}) are not suitable as a starting point. The
reason is that in subcritical case constraints (\ref{quantconst})
have different signs in left and right modes, see Eq.
(\ref{chir+-sub}) for $\varepsilon_+=\varepsilon_-=1$. As a
consequence, there is no linear combination of these complex
constraints  which in the flat limit would give two independent
constraint equations. To find such a suitable set of complex
constraints, it is more convenient to proceed directly from
non-chiral classical constraints (\ref{constNon1}) and
(\ref{constNon2}). For a sake of definiteness we put $\kappa>0$,
and so, $\varepsilon_+=\varepsilon_-=1$. Linear combination
$\Pi_1+i\Pi_2$ at the quantum level has a nontrivial kernel, and
can be chosen as one of the sought for quantum constraint
equations,
\begin{equation}\label{Pi+Phys}
    (\Pi_1+i\Pi_2)\Psi=0.
\end{equation}
To fix another constraint, let us take a linear combination of $V_i$
and $\Pi_i$,
\begin{equation}
    \tilde{V}_{i}=V_{i}-\frac{m}{\kappa }\epsilon _{ij}\Pi _{j}=
    p_{i}+\frac{%
\kappa }{2R^{2}}\epsilon _{ij}x_{j}-\frac{m}{\kappa }\epsilon
_{ij}\left( \pi _{j}-\frac{\kappa }{2}\epsilon _{jk}v_{k}\right),
\end{equation}
 which is decoupled from the constraints $\Pi_i$ in the
sense of brackets,
\begin{eqnarray}
    &\{ \tilde{V}_{i},\Pi _{j}\} =0,\label{tilViPi}&\\
    &\{ \tilde{V}_{i},\tilde{V}_{j}\} =-\frac{m^2}{\kappa}(1-\rho^2)\epsilon_{ij}.\label{tilVitilVi}&
\end{eqnarray}
Relations (\ref{tilViPi}), (\ref{tilVitilVi}) mean that linear
combination $\tilde{V}_i$ is a Dirac extension of $V_i$ with respect
to the second class constraints $\Pi_i\approx 0$ [cf. Eq.
(\ref{tilVitilVi}) with Dirac brackets (\ref{VV})]. According to
(\ref{tilVitilVi}), at the quantum level linear combination
$\tilde{V}_1-i\tilde{V}_2$ has a nontrivial kernel (it is of a
nature of annihilation operator), and quantum constraint
(\ref{Pi+Phys}) can be supplied with the wave equation
\begin{equation}\label{Til-Phys}
    (\tilde{V}_1-i\tilde{V}_2)\Psi=0.
\end{equation}

Eqs. (\ref{NCconsistency}) give the following  Poisson bracket
relations for the total Hamiltonian (\ref{HtotNC}) with the chosen
combinations of the constraints
\begin{eqnarray}
    \{H,\Pi_1+i\Pi_2\}&=&-i\frac{m}{\kappa}(\Pi_1+i\Pi_2)+(\tilde{V}_1+i\tilde{V}_2),\label{HtotViPi1}\\
    \{H,\tilde{V}_1-i\tilde{V}_2\}&=&-i\frac{m}{\kappa}(\tilde{V}_1-i\tilde{V}_2)+\frac{m^2}{\kappa^2}
    (1-\rho^2)(\Pi_1-i\Pi_2).\label{HtotViPi2}
\end{eqnarray}
On the right hand side, there appear complex conjugate combinations
of the chosen constraints. This means that the quantum dynamics
generated by the Schr\"odinger equation with the total Hamiltonian
taken as a Hamiltonian operator would be not consistent with quantum
constraints (\ref{Pi+Phys}), (\ref{Til-Phys}). One can overcome this
obstacle if we pass from the total Hamiltonian to the corrected one,
$H\rightarrow \tilde H$, by adding the terms quadratic in the
constraints,
\begin{equation}\label{H*}
    \tilde H=H-\frac{1}{\kappa}\epsilon_{ij}\Pi_i\tilde{V}_j+
    \frac{m}{2\kappa^2}(\Pi_1-i\Pi_2)(\Pi_1+i\Pi_2)+\frac{1}{2m(1-\rho^2)}(\tilde{V}_1+i\tilde{V}_2)
    (\tilde{V}_1-i\tilde{V}_2).
\end{equation}
The corrected Hamiltonian commutes strongly with all the constraints
$\Pi_i$ and $\tilde{V}_i$, and, in particular,   with the chosen
complex linear combinations of the constraints. Then the
Schr\"odinger equation
\begin{equation}\label{SchH*}
    \left(i\partial_t-\tilde H\right)\Psi=0
\end{equation}
will be consistent with quantum constraints (\ref{Pi+Phys}),
(\ref{Til-Phys}): physical state satisfying the quantum constraint
equations at $t=0$ will  also satisfy them for any $t>0$. Note
that due to commutativity of the corrected Hamiltonian with the
constraints, it is a Dirac extension of the canonical Hamiltonian
with respect to the set of all the four second class constraints.

We do not display an explicit form of the physical states satisfying
quantum equations (\ref{Pi+Phys}), (\ref{Til-Phys}), but instead let
us discuss the flat limit of the system. For $R\rightarrow \infty$,
Lagrangian (\ref{nhaction}) reduces to Lagrangian of a free particle
on the non-commutative plane,
\begin{equation}\label{Lflat}
{\cal L}_{nc}=
    mv_i\dot x_i-m\frac{v_i^2}{2}
    +\frac 12\kappa\epsilon_{ij}v_i\dot v_j.
\end{equation}
System (\ref{Lflat}) has  an exotic Galilei symmetry given by Eqs.
(\ref{NH3ex})---(\ref{PP}) with $R=\infty$. On the reduced phase
space the system is described by a symplectic structure with
noncommutative coordinates,
\begin{equation}\label{xxpfree}
    \{x_i,x_j\}=\frac{\kappa}{m^2}\epsilon_{ij},\qquad
    \{x_i,p_j\}=\delta_{ij},\qquad \{p_i,p_j\}=0,
\end{equation}
 and by a free
Hamiltonian
\begin{equation}\label{H*free}
    H^*=\frac{1}{2m}p_i^2,
\end{equation}
which are the flat limits of  (\ref{xxpx}) and  (\ref{Hredxp}).

 On the other hand, constraints (\ref{Pi+Phys}),
(\ref{Til-Phys}) and Schr\"odinger equation (\ref{SchH*}) in the flat
limit reduce, respectively, to
\begin{equation}\label{ef1}
\left( \frac{\partial }{\partial v_{-}}+\frac{\kappa
}{4}v_{+}\right) \Psi \left( x,v,t\right) =0,
\end{equation}
\begin{equation}\label{ef2}
\left( \frac{\partial }{\partial x_{+}}-i\frac{m}{4}v_{-}-i\frac{m}{\kappa }%
\frac{\partial }{\partial v_{+}}\right) \Psi \left( x,v,t\right) =0,
\end{equation}
\begin{equation}\label{ef3}
\left( i\frac{\partial }{\partial t}+\frac{2}{m}\frac{\partial ^{2}}{%
\partial x_{+}\partial x_{-}}\right) \Psi \left( x,v,t\right) =0,
\end{equation}
where we use the notations $x_\pm=x_1\pm ix_2$, $v_\pm=v_1\pm iv_2$.
General solution to the constraint equations is
\begin{equation}\label{Psiphys2}
    \Psi_{phys}=\exp\left(-\frac{\kappa}{4}v_+v_-\right)
    \psi\left(x_+-i\frac{\kappa}{m}v_+,x_-,t\right),
\end{equation}
and a general solution to the Schr\"odinger equation being an
eigenstate of the momentum operator with eigenvalue $\vec{p}$ can
be presented in the form
\begin{eqnarray}
    \Psi_{\vec{p}}\,(\vec{x},\vec{v},t)&=&\exp\left(-\frac{\kappa}{4}\left(v_+v_-+\frac{\vec{p}\,{}^2}{m^2}
    \right)
    -i\frac{\vec{p}\,^2}{2m}t+i\vec{p}\vec{x}
    +\frac{1}{2}\frac{\kappa}{m}p_- v_+\right),\label{Psiop1}\\
    &=&\exp\left(-\frac{\kappa}{4}\left(\vec{v}-\frac{1}{m}\vec{p}\right)^2
    +\frac{i}{2}\frac{\kappa}{m}\epsilon_{ij}p_iv_j
    -i\frac{\vec{p}\,^2}{2m}t+i\vec{p}\vec{x}\right),\label{Psiop2}
\end{eqnarray}
where $p_-=p_1-ip_2$. A specific dependence on $v_i$ guarantees a
normalizability of the state (\ref{Psiop1}) in velocity variables.
Note that Eq. (\ref{Pi+Phys}) can be presented in the form
$(a_1-a_2)\Psi=0$, where $a_{1,2}$ are the annihilation operators
constructed in terms of $v_{1,2}$ and their derivatives. Hence, this
equation means that physical states correspond to $n_1=n_2$, where
$n_1$, $n_2$ are the eigenvalues of the number operators
$N_1=a^\dagger_1a_1$ and $N_2=a^\dagger_2a_2$, i.e. physical states
have a definite ``circular polarization" in velocity variables. At
$\kappa=0$ wave function (\ref{Psiop2}) turns into a planar wave of
an ordinary free planar particle.

In  conclusion of this section we find a relation of the state
(\ref{Psiop2}) with wave function of the free exotic particle in
Fock space representation \cite{PHMP}. In Fock space representation
approach, we first eliminate at the classical level momenta $\pi_i$
and then quantize proceeding from the symplectic structure given by
Eq. (\ref{xpvv}). The physical subspace is given in this case by the
equation
\begin{equation}\label{physFock}
\left(-2i\frac{\partial}{\partial x_+}-m\hat v_-\right)\Psi=0,
\end{equation}
where $[\hat v_-,\hat v_+]=\frac{2}{\kappa}$. We realize $\hat
v_\pm$ by oscillator operators, $\hat v_-=\sqrt{2\kappa^{-1}}\,a$,
$\hat v_+=(v_-)^\dagger$, $[a,a^\dagger]=1$, decompose the state in
terms of velocity number operator eigenstates,
\begin{equation}\label{sum0inf}
    \Psi=\sum_{n=0}^{\infty}\phi_n(x)\vert n\rangle,
\end{equation}
and find that for physical states all the components $\phi_n$ with
$n>1$ can be represented in terms of $\phi_0$,
\begin{equation}\label{phin0}
    \phi_n=(-1)^n\left(\frac{\kappa}{2}\right)^{\frac{n}{2}}
    \frac{1}{\sqrt{n!}}\left(\frac{\hat p_-}{m}\right)^n\phi_0,
\end{equation}
where $\hat p_-=-2i\partial/\partial x_+$, see Eq. (3.6) in
\cite{PHMP}. Having in mind a correspondence between the Fock
space and holomorphic representations,
\begin{equation}\label{Fockhol1}
    a^\dagger\leftrightarrow z,\quad  a\leftrightarrow
    \frac{d}{dz},\quad
    \vert n\rangle \leftrightarrow \exp\left(-\frac{1}{2}\vert z\vert^2\right)
    \frac{z^n}{\sqrt{n!}},
\end{equation}
where $z$  is a complex variable, we identify
$v_+=\sqrt{\frac{2}{\kappa}}\,z$, and decompose
\begin{equation}\label{coherent}
    \exp\left(\frac{1}{2}\frac{\kappa}{m}p_- v_+\right)
\end{equation}
in the Taylor series. As a result we find that physical state
(\ref{Psiop1}) is equivalent to the state (\ref{sum0inf}) with
$\phi_n$ given by  (\ref{phin0}). Note that in accordance with Eq.
(\ref{physFock}),  the state (\ref{sum0inf}), (\ref{phin0}) is a
coherent state of the velocity annihilation operator $\hat v_-$
with operator-valued  eigenvalue $\frac{1}{m}\hat p_-$, and  in
the wave function (\ref{Psiop1}) it is the factor (\ref{coherent})
that reflects such a nature of a physical state.

\section{Projective phase and two-cocycle}\label{Phaseproj}

Here we compute a projective phase corresponding to the exotic
NH$_3$ symmetry.  This phase is related to the non-trivial
two-cocycle of the exotic NH$_3$ group. The two-cocycle can be
computed by direct calculation, or from the quasi-invariance of our
particle Lagrangian under translations and boosts
\cite{bargmann,levyleblond69,azcarragabook}. Its presence guarantees
the invariance of the wave equations.

Consider a unitary operator
\begin{equation}\label{Utrbo}
    U(\alpha,\beta)=\exp i(\alpha_iP_i-\beta_iK_i)
\end{equation}
in the  non-chiral basis, where $P_i$ and $K_i$ are implied to be
quantum analogs of the classical integrals (\ref{PKNCobs}). Using
the commutation relations (\ref{KPK}) and (\ref{PP}) and $ e^{A}e^B=
e^{A+B}e^{\frac{1}{2}[A,B]},$ valid for any operators $A$ and $B$
satisfying a relation $[A,[A,B]]=[B,[A,B]]=0,$ we obtain the
following composition law
\begin{equation}\label{UUaabb}
    U(\alpha,\beta)U(\alpha',\beta')=e^{-i\omega_2(\alpha,\beta;\alpha',\beta')}
    U(\alpha+\alpha',\beta+\beta'),
\end{equation}
where  a nontrivial phase factor is equal (modulo $2\pi n$, $ n\in
\Z$) to
\begin{equation}\label{omega2}
    -\omega_2(\alpha,\beta;\alpha',\beta')=-\frac{1}{2}\left(\kappa\epsilon_{ij}
    \left(\frac{1}{R^2}\alpha_i\alpha'_j+\beta_i\beta'_j\right)
    +m(\alpha_i\beta'_i-\beta_i\alpha'_i)\right).
\end{equation}
A direct calculation shows that (\ref{omega2}) satisfies a  zero
coboundary condition,
\begin{equation}\label{}
    \Delta \omega_2\equiv
    \omega_2(g_2,g_3)-\omega_2(g_1g_2,g_3)+\omega_2(g_1,g_2g_3)-\omega_2(g_1,g_2)=0,
\end{equation}
which guarantees the associativity of the product (\ref{UUaabb}).
Here $g_{1}$, $g_2$  mean group elements, which in our case are
characterized by the sets of parameters $(\alpha_i,\beta_i)$,
$(\alpha'_i,\beta'_i)$, with a composition law $g_1g_2\rightarrow
(\alpha_i+\alpha'_i,\beta_i+\beta'_i)$. Therefore
$\omega_2(\alpha,\beta;\alpha',\beta')$ is the two-cocycle of NH$_3$
group.

Now let us consider the action of the unitary operator (\ref{Utrbo})
on coordinates $x_i, v_i$.  In correspondence with  (\ref{trbo1}),
(\ref{trbo2}), it generates the NH translation and boost
transformations,
\begin{equation}\label{xxtrbo}
    x'_i=Ux_iU^{-1}=x_i+ R {\cal A}_i(t),\qquad
    v'_i=Uv_iU^{-1}=v_i+{\cal B}_i(t),
\end{equation}
where we have introduced a compact notation for a rotation in a
`plane' $(\frac{1}{R}\alpha_i,\beta_i)$ of dimensionless parameters,
\begin{equation}\label{ABalbeta}
    {\cal A}_i(t)=\frac{1}{R}\alpha_i\cos R^{-1}t + \beta_i\sin
    R^{-1}t,\qquad
    {\cal B}_i(t)=\beta_i\cos R^{-1}t - \frac{1}{R}\alpha_i\sin
    R^{-1}t,
\end{equation}
 ${\cal A}_i(0)=\frac{1}{R}\alpha_i$,  ${\cal B}_i(0)=\beta_i$.
 In terms of (\ref{ABalbeta}), we have
\begin{equation}\label{PKAB}
    \alpha_iP_i-\beta_iK_i= A(x,v;\alpha,\beta) +B(p,\pi;\alpha,\beta),
\end{equation}
where we have separated the coordinate and momenta depending parts,
\begin{eqnarray}
    A(x,v;\alpha,\beta)&=&-m{\cal
    B}_i(t)x_i-\frac{\kappa}{2}\epsilon_{ij}\left(R^{-1}{\cal
    A}_i(t)x_j+{\cal B}_i(t)v_j\right),\nn\\
    B(p,\pi;\alpha,\beta)&=&R{\cal A}_i(t)p_i+{\cal B}_i(t)\pi_i.\label{ABxvt}
\end{eqnarray}
As a result, the action of (\ref{Utrbo}) on a wave function can be
represented in a form
\begin{equation}\label{Upsixv}
    \tilde{\Psi}(x,v,t)\equiv
    U(\alpha,\beta)\Psi(x,v,t)=e^{-i\omega_1(x,v,t;\alpha,\beta)}\Psi(x',v',t),
\end{equation}
where $x'$ and $v'$ are given by Eq. (\ref{xxtrbo}), and a phase
$\omega_1$ is given (modulo $2\pi n$, $n\in \Z$)  by
\begin{equation}\label{omega_1F}
    \omega_1(x,v,t;\alpha,\beta)=m{\cal
    B}_i(t)\left(x_i+\frac{1}{2}R{\cal A}_i(t)\right)
    +\frac{\kappa}{2R}\epsilon_{ij}\left({\cal A}_i(t)x_j+R{\cal
    B}_i(t)v_j\right).
\end{equation}
$\omega_1(x,v,t;\alpha,\beta)$ is the real-valued one-cocycle,
projective phase, of NH$_3$.

From (\ref{UUaabb}) and  (\ref{Upsixv})  we can see
 the two-cocycle (\ref{omega2})  can be
 written in terms of the projective phase (\ref{omega_1F}),
\begin{equation}\label{ome1om2}
    \omega_2(g_1,g_2)=\Delta\omega_1\equiv \omega_1(q^{g_1};g_2)+\omega_1(q;g_1)-\omega_1(q;g_1g_2),
\end{equation}
where $q$ means the set $(x_i,v_i,t)$, and  $q^g=gq$ means an
application of $g$ to coordinates $q$, that in our case corresponds
to  $(x'_i,v'_i,t')$ with $t'=t$ and transformed coordinates
(\ref{xxtrbo}).

The  projective phase is also associated with quasi-invariance of
Lagrangian with respect to corresponding classical symmetry
transformations, see \cite{azcarragabook}.
A direct check shows that
in our case under classical symmetry transformation of the form
(\ref{xxtrbo}), the nonchiral Lagrangian (\ref{nhaction}) transforms
as
\begin{equation}\label{omFL}
    {\cal L}'_{nc}={\cal L}_{nc}+\delta {\cal L}_{nc},\qquad
    \delta {\cal L}_{nc}=\frac{d}{dt}\left(\omega_1(x,v,t;\alpha,\beta)\right).
\end{equation}
\vskip0.3cm

In the flat limit $R\rightarrow \infty$  (\ref{omega_1F}) reduces to
\begin{equation}\label{Fflat}
    \omega_1^0(x,v,t;\alpha,\beta)=m\beta_ix_i+\frac{1}{2}m\beta_i^2t+\frac{\kappa}{2}\epsilon_{ij}\beta_iv_j.
\end{equation}
The 2-cocycle in this case is
\begin{equation}\label{omega20}
    -\omega_2^0(\alpha,\beta;\alpha',\beta')=-\frac{1}{2}\left(\kappa\epsilon_{ij}
    \beta_i\beta'_j
    +m(\alpha_i\beta'_i-\beta_i\alpha'_i)\right).
\end{equation}

\vskip0.3cm

In the same way  one can compute a projective phase in the chiral
basis. It appears under the action of the (chiral) analog of the
unitary operator (\ref{Utrbo}) on a chiral wave function,
\begin{equation}\label{Upsixv+-}
    \tilde{\Psi}(X^+,X^-,t)\equiv
    U_{ch}(\alpha,\beta)\Psi(X^+,X^-,t)=e^{-i\tilde{\omega}_1(X^+,X^-,t;\alpha,\beta)}\Psi(X^+{}',{X^-}{}',t),
\end{equation}
\begin{equation}\label{FaseChir2}
    \tilde{\omega}_1(X^+,X^-,t;\alpha,\beta)=-\frac{1}{2}\mu_+\epsilon_{ij}X^+_i\alpha^+_j(t)+
    \frac{1}{2}\mu_-\epsilon_{ij}X^-_i\alpha^-_j(t).
\end{equation}
Here $\alpha^\pm_i(t)$ are defined  by Eq. \bref{albet+-} and
the translation and boost generators are constructed according to
(\ref{PKchiral}). The transformation \bref{xxtrbo} is changed in the chiral
variables as in \bref{NHchiral*}. As in the non-chiral case,  for the chiral
formulation the time-dependent shift symmetry of the coordinates
under translation and boost transformations \bref{NHchiral*}
produces a change in the chiral Lagrangian \bref{Lchir}, which is
given by the projective phase
\begin{equation}\label{Lchir3}
    {\cal L}'_{ch}={\cal L}_{ch} +
    \frac{d}{dt}\tilde{\omega}_1(X^+,X^-,t;\alpha,\beta).
\end{equation}

A computation of the two-cocycle on the basis of the
Baker-Campbell-Hausdorff formula gives the same result
as in the non-chiral formulation  \bref{omega2}
since  it is based on the same
exotic Newton-Hooke algebra, and in particular, on the same
commutation relations \bref{KPK} and \bref{PP}. The identity of the
two-cocycles in both formulations can also be understood from the
point of view of the canonical transformation associated with a
total time derivative difference (\ref{Nc-Ch}) between the two forms
of Lagrangians. Indeed, in correspondence with classical relation
(\ref{Nc-Ch}), the projective phases in chiral and non-chiral
formulations are related as
\begin{equation}\label{phasechirnonchir}
    \tilde{\omega}_1-\omega_1\equiv
     \rho(x,v,t;\alpha,\beta)
    =-\frac{1}{2}m(x'_iv'_i-x_iv_i)=
     \frac{mR}{2}\left({\cal
    A}_i(t)v_i+{\cal B}_i(t)x_i+{\cal A}_i(t){\cal
    B}_i(t)\right),
\end{equation}
where $x'_i$ and $v'_i$ are given in Eq. \bref{xxtrbo}. The
difference is a real valued trivial one-cocycle of the form
$f(q^g)-f(q)$.

Finally, let us note that canonical transformation \bref{chi-nochi}
is behind the following relation between the chiral and non-chiral
forms of the unitary operator \bref{Utrbo}, that represents NH
translation and boost transformations,
\begin{equation}\label{UchirUnc}
 {U}_{ch}=e^{-i\rho(x,v,t;\alpha,\beta)}U_{nc},
\end{equation}
where $\rho(x,v,t;\alpha,\beta)$ is defined in
\bref{phasechirnonchir}.

\vskip 0.3cm
 Let us consider now the
covariance of the wave equations we have introduced. They can be
written as
\begin{equation}\label{covEq1}
    {\cal D}_a\Psi(q,t)=0,
\end{equation}
\begin{equation}
    \left(i\partial_t -H\right)\Psi(q,t)=0,\label{covEq2}
\end{equation}
where $q$ denotes coordinates $x_i,\, v_i$  or $X^+_i,\, X^-_i$ for
the  non-chiral or chiral formulation, \bref{covEq1} is a set of two
quantum constraints, whose form depends on the phase we consider,
and quantum Hamiltonian $H$ is adjusted with them. The invariance of
the theory under time translations is obvious, and since for any
quantum constraint $ [J,{\cal D}_a]\propto {\cal D}_a, $ the
rotation invariance is  obvious too. Analogous conclusion is valid
for additional symmetry associated with integrals $I_1,$ $I_2$.
Futher, since operators $P_i$ and $K_i$ commute with quantum
constraints, wave equations \bref{covEq1} are invariant under NH
translations and boosts transformations. Finally, in correspondence
with a classical relation $\partial \Gamma_i/\partial
t+\{\Gamma_i,H\}=0$, $\Gamma_i=P_i,K_i$, at the quantum level
operator  \bref{Utrbo} commutes with the operator $i\partial_t-H$.
Therefore, we have
\begin{equation}
    (i\partial_t-H)\Psi(q,t)=0,\quad\to\quad
    (i\partial_t-H)U\Psi(q,t)=(i\partial_t-H)e^{-i\omega_1(q,t;g)}\Psi(q^g,t)=0,
\end{equation}
i.e.  the transformed wave function \bref{Upsixv} or \bref{Upsixv+-}
becomes the solution of the Schr\"odinger equation.

Summing up, the projective phase guarantees the covariance of our
wave equations.

\section{Discussion and outlook}\label{discussionoutlook}

Duality transformation (\ref{duality3}),
$\rho=\frac{\kappa}{mR}\rightarrow \rho^{-1}$, which relates sub-
and super-critical phases of the model, can be reinterpreted as a
kind of $T$-duality,
\begin{equation}\label{Tdual}
R\rightarrow \left(\frac{\kappa}{m}\right)^2\frac{1}{R}\,.
\end{equation}
Then the exotic Newton-Hooke particle in the sub- (super-)critical
phase characterized by the parameters  $m$, $\kappa$ and $R$ with
$\rho^2<1$ ($\rho^2>1$), can be related in a unique way to the same
system  in the super- (sub-)critical phase with inverse value of the
radius parameter given by Eq. (\ref{Tdual}). Such a correspondence
implies the dual relation (\ref{JHdual}) between Hamiltonian and
angular momentum of the both systems, as well as the duality
relation  (\ref{PBo1})  between Newton-Hooke translations and
boosts, or the duality  (\ref{chirJJJ}) for the chiral generators.
It also implies the change of associated additional symmetry, that
defines the degree of degeneration of energy levels, from the
compact, so(3) ($\rho^2<1$), to the non-compact, so(2,1)
($\rho^2>1$), one.

In our system, the duality relates different sectors  corresponding
to different values of the model parameters. The parameters $m$
and $\kappa$ can be promoted to be dynamical variables if we treat
them in Lagrangian (\ref{nhactionZZ}) as momenta canonically
conjugate to variables $c$ and $\tilde c$. As a result,  different
phases of the model will be realized in different parts of the
extended phase space. Note that this phenomenon is similar to that
observed earlier in  Lovelock gravity \cite{TeiZan} and higher
dimensional pure Chern-Simons theories \cite{BanHen}.

We analyzed the `trigonometric' (periodic) case of the (2+1)D exotic
Newton-Hooke symmetry. The results can be translated in a simple way
for the `hyperbolic' case of the exotic NH$_3$, which appears under
contraction of dS$_3$ \cite{Mariano,Gao}. The dS$_3$ algebra follows
from AdS$_3$ algebra (\ref{P0M12})--((\ref{PiPj}) via a simple
substitution $R^2\rightarrow -R^2$. A corresponding Lagrangian can
be obtained from our non-chiral Lagrangian (\ref{nhaction}) via the
same substitution.   As a result, in analog of relation (\ref{detA})
that characterizes the algebra of constraints, the quantity
$m^4(1-\rho^2)^2$ will be changed for $m^4(1+\rho^2)^2$. Therefore,
in hyperbolic case the constraints form the set of second class
constraints for any choice of the parameters $m$ and $\kappa$, and
corresponding exotic Newton-Hooke system has only one phase.

The chiral Lagrangian (\ref{Lchir}) with the changed signs before
the potential terms for both chiral modes corresponds to such a
case. Since the existence of different phases is rooted in the
properties of the constraints, which all are the primary
constraints defined by a kinetic part of Lagrangian, the
hyperbolic case is characterized by the same set of phases related
by a duality transformation.

To conclude, let us list some open problems to be interesting for
further investigation.

We analyzed  here the exotic Newton-Hook symmetry in 2+1 dimensions
proceeding from non-relativistic contraction of AdS$_3$. The reduced
phase space description of the  model reveals a symplectic structure
similar to that of Landau problem in a non-commutative plane. The
latter system, as a present one, also reveals sub- and
super-critical phases separated by a critical phase. Therefore, it
would be natural to investigate the noncommutative Landau problem
\cite{HP3} in the light of the exotic Newton-Hooke symmetry
\cite{InPre}.

One could expect that if we construct a Lagrangian for a
relativistic particle on AdS$_3$ space by the method of nonlinear
realization, in appropriate non-relativistic limit it should
reduce to the model investigated here. Therefore, the interesting
question is what would correspond in a relativistic model to the
present sub-, super- and critical phases, and what would be the
analog of the duality transformation there?

It would be interesting  to generalize the exotic Newton-Hooke
particle model for the supersymmetric case. Since in our bosonic
model in super-critical case Hamiltonian is not positively
definite, one could expect the appearance of some restrictions on
the domain of the model parameters  in the context of
supersymmetric extension.

There are some indications  that the investigated model should have
a close relation to the physics of  BTZ black hole.  Indeed, the
AdS$_3$ structure underlies the BTZ black hole
\cite{Banados:1992wn}, which also reveals different phases in
dependence on the values of its mass and angular momentum. The
chiral form of our Lagrangian (\ref{Lchir}) is reminiscent of the
Lagrangian in Chern-Simons formulation of 3D gravity
\cite{AchTown,Witten}. If such a relation really exists, it would be
interesting to clarify, in particular,  what in BTZ black hole
physics should correspond to the duality of the exotic Newton-Hooke
particle system.

\vskip 0.4cm\noindent {\bf Acknowledgements}. We are grateful to
Jorge Alfaro, Luis Alvarez-Gaume, Adolfo Azcarraga, Max Ba\~nados,
Mokhtar Hassaine, Mariano del Olmo, Dimitri Sorokin,  Paul Townsend,
Toine Van Proeyen and Jorge Zanelli for stimulating discussions.  MP
thanks the Physics Department of Barcelona University, where a part
of this work was realized, for hospitality. JG thanks the Physics
Department of Universidad de Santiago de Chile for hospitality. The
work was supported in part by CONICYT, FONDECYT Project 1050001,
MECESUP USA0108,   the European EC-RTN network MRTN-CT-2004-005104,
MCYT FPA 2004-04582-C02-01 and CIRIT GC 2005SGR-00564.

\section{Appendix }\label{AppendixA}

In this appendix we  compute the non-trivial Eilenberg-Chevalley
cohomology for the Newton-Hooke group in 2+1 dimensions. The
unextended Newton-Hooke algebra \cite{BacLev,BacNuy} is given by
\begin{equation}\label{NH3ex*}
    [H,J]=0,
\end{equation}
\begin{equation}\label{HPK*}
    [H,K_i]=-iP_i,\qquad [H,P_i]=i\frac{1}{R^2}K_i,
\end{equation}
\begin{equation}\label{JKP*}
    [J,P_i]=i\epsilon_{ij}P_j,\qquad
    [J,K_i]=i\epsilon_{ij}K_j,
\end{equation}
\begin{equation}\label{KPK*}
    [K_i,P_j]=
    [K_i,K_j]=0,
\end{equation}
\begin{equation}\label{PP*}
    [P_i,P_j]=0.
\end{equation}
Consider a group element
\begin{equation}
\label{coset1*} g=e^{-iH
x^0}\;e^{iP_ix^i}\;e^{iK_jv^j}e^{-iJ\theta},
\end{equation}
where $\theta$ is a local coordinate on $S^1$.  The Maurer-Cartan
one-form is given by
 \begin{equation}\label{MCar*}
    \Omega=-ig^{-1}dg= -L_H H +L^i_P P_i+L^i_K K_i+L_J J,
\end{equation}
where
\begin{eqnarray}
    L_H &=& dx^0,
    \qquad
    L^i_P ={ R^i}_j(\theta)(dx^j-v^j dx^0),
    \qquad
    L^i_K = {R^i}_j(\theta) \left(dv^j+\frac{x^j}{R^2}\,dx^0\right),
    \qquad
    L_J = -d\theta.
\nn\\
\end{eqnarray}
with ${R^i}_j(\theta)=\left(\begin{array}{cc} \cos
\T&\sin\T\\-\sin\T&\cos\T
\end{array}\right)$
  being an SO(2) rotation.

 There are two closed
rotation-invariant two-forms,
\begin{eqnarray}
\Omega_2=L^i_K \wedge L^i_P,\qquad d\Omega_2=0,
\end{eqnarray}
\begin{eqnarray}
\tilde{\Omega}_2=\frac 12\epsilon_{ij}\left[L^i_K\wedge L^j_K+
\frac{1}{R^2}L^i_P \wedge L^j_P\right],\qquad d\tilde{\Omega_2}=0.
\end{eqnarray}
They are expressed locally as
\begin{equation}
\Omega_2=d\Omega_1,\qquad
\Omega_1=v^idx^i-\frac{v_i^2}{2}dx^0-\frac{x_i^2}{2R^2}\,dx^0,
\end{equation}
\begin{equation}
\tilde{\Omega}_2=d\tilde\Omega_1,\qquad
    \tilde\Omega_1=\frac{1}{2}\epsilon_{ij}
    \left(v^idv^j+\frac{1}{R^2}x^i dx^j-
    \frac{2}{R^2}x^iv^jdx^0\right).
\end{equation}
The one-forms $\Omega_1$ and  $\tilde\Omega_1$ are not
left-invariant.

There is also a third closed rotation-invariant form,
\begin{equation}\label{Om3}
\check{\Omega }_{2}=L_{H}\wedge L_{J},\qquad d\check{\Omega }_{2}=0,
\end{equation}
which is expressed locally as
\begin{equation}\label{Om3*}
    \check{\Omega }_{2}=d\check{\Omega }_{1},\qquad
    \check{\Omega }_{1}=\theta dx^{0}.
\end{equation}
The one-form $ \check{\Omega }_{1}$ is not left-invariant either.
Therefore the Eilenberg-Chevalley cohomology of degree 2 is
non-trivial. This implies that the Newton-Hooke algebra has a
three-fold central extension
\begin{equation}\label{HJg}
    \left[ H,J\right] =\check{Z}.
\end{equation}
\begin{equation}
    \lbrack K_{i},P_{j}]=i\delta _{ij}Z,\qquad \lbrack
    K_{i},K_{j}]=-i\epsilon _{ij}\tilde{Z},  \label{KPK**}
\end{equation}%
\begin{equation}
    \lbrack P_{i},P_{j}]=-i\frac{1}{{R}^{2}}\epsilon _{ij}\tilde{Z},
\label{PP**}
\end{equation}
However, we neglect the extension associated with the central
element $\check{Z}$.  Such a three-fold extension of NH$_3$ algebra
cannot be obtained by a contraction of AdS$_3$, cf.  Eq.
(\ref{P0M12}) and (\ref{HJg}). In the presence of the third central
element $\check Z$, the unique Casimirs of the algebra are the three
central elements, and so, $H$ and $J$ cannot be presented in terms
of the boosts and translations generators, see \cite{BGGK,Gao}. The
third algebra extension does not give any non-trivial contribution
to our exotic particle Lagrangian \cite{FootApp}. Therefore, we put
$\check Z=0$.

Then, another implication of the non-trivial cohomology is that we
have two Wess-Zumino terms,
\begin{equation}
S_1=\int \Omega_1^*,\qquad \tilde S_1=\int \tilde\Omega_1^*,
\end{equation}
whose linear combination describes the exotic particle dynamics;
here $*$ means a pullback on the world-line of the particle.



\begin{thebibliography}{99}
\bibitem{FootApp}
If we add an extra exponential with new central extension in Eq
(\ref{coset1}), the Maurer-Cartan form gets a new term  $dy$, where
$y$ is the corresponding Goldstone field associated to the third
central charge. The contribution of this term to the Lagrangian is a
total derivative.
\end{thebibliography}
\end{document}